\documentclass[fleqn,10pt]{olplainarticle}
\usepackage{subcaption}
\usepackage{url}
\usepackage{xcolor}      % for \textcolor in \edo{}
\usepackage{booktabs}    % for \toprule, \midrule, \bottomrule in tables

\usepackage{enumitem}    % for [noitemsep] option in itemize/enumerate

\newcommand{\SFig}[1]{Figure~\ref{#1}}
\newcommand{\STab}[1]{Table~\ref{#1}}

\title{Twitter climate discourse as a signal of pro-environmental behaviors}

\author[1,2,3]{Edoardo Maggioni}
\author[2,4]{Diego Garlaschelli}
\author[5]{Rossana Mastrandrea}
\author[6]{Luca Maria Aiello}
\affil[1]{Department of Computer Science, University of Pisa, Pisa, Italy}
\affil[2]{IMT School for Advanced Studies, Lucca, Italy}
\affil[3]{Institute of Informatics and Telematics (IIT), National Research Council (CNR), Pisa, Italy}
\affil[4]{Lorentz Institute for Theoretical Physics, University of Leiden, Leiden, Netherlands}
\affil[5]{Department of Management "Valter Cantino", University of Turin, Turin, Italy}
\affil[6]{Data Science Section, IT University of Copenhagen, Copenhagen, Denmark}


\begin{abstract}

Fostering coordinated pro-environmental behaviors at scale is a key challenge for climate mitigation. Individual actions only generate meaningful impact when they diffuse widely and become socially coordinated, yet monitoring such processes remains difficult with traditional survey-based tools alone.

In this study, we examine whether large-scale online climate discourse is associated with differences in offline pro-environmental behavior across European regions. We combine geolocated Twitter data from the Climate Change Twitter Dataset (2017--2019) with survey-based measures from the 2019 Special Eurobarometer, focusing on the regional density of climate-related tweets and the average number of self-reported pro-environmental actions.

We find a strong positive association between tweet density and pro-environmental behavior that remains robust to socio-economic controls, alternative spatial aggregations, and a wide range of robustness checks. To move beyond aggregate volume, we further decompose online discourse using Natural Language Processing tools that capture distinct social dimensions. While knowledge exchange shows no clear relationship with offline behavior, the prevalence of activism- and social support--related expressions is negatively associated with pro-environmental actions.

Overall, our results suggest that online climate discourse can serve as an informative, attention-related signal of regional differences in pro-environmental behavior, but that different forms of online engagement relate to offline action in markedly different ways. More broadly, the study highlights the potential of integrating large-scale digital traces with survey data to investigate collective behavior in socio-environmental systems, while remaining explicitly observational in scope.
\end{abstract}

\begin{document}

\flushbottom
\maketitle
\thispagestyle{empty}

%\noindent\textbf{Disclaimer:} This manuscript is a working paper in its final stage. It has not yet undergone full revision and approval by all co-authors and may differ from the version submitted for publication.

\section{Introduction}

% SITUATION
Environmental change is one of the major challenges of our time, and a wide range of policies have been introduced to reduce the negative impact of human activity on the climate and the environment. Over the past decades, international agreements such as the Kyoto Protocol and the Paris Agreement have pushed many developed countries, and the European Union in particular, to adopt measures aimed at reducing CO$_2$ emissions and limiting pollution. However, these top-down interventions alone are not sufficient to substantially alter current trajectories: their effectiveness ultimately depends on whether they are adopted and translated into concrete actions by the general public.
This process operates at multiple levels, involving both large-scale actors, such as industries, energy producers, and data centers, and individuals through their everyday decisions.

At the individual level, pro-environmental behaviors related to energy use, mobility, and consumption patterns play a crucial role. Yet, their impact emerges only when these actions become widespread and reach a collective scale. Understanding how pro-environmental behaviors diffuse and become socially coordinated is therefore important both from a theoretical perspective and for policy design. At the same time, it requires tools that allow such behaviors to be monitored at scale and over time, beyond what traditional survey-based approaches can typically provide.

% COMPLICATION
Social media appear to offer a promising source of such information. They generate continuous, fine-grained data and often provide geolocation signals that can be exploited at large scale. As a result, a growing body of work uses social media data to study opinions, attention, sentiment, and online activism around environmental issues \citep{pearce2019social,sultana2024systematic}. However, linking these online expressions to offline behavior remains challenging. Most existing studies focus exclusively on online dynamics, and only a limited number explicitly relate online discourse to independently measured offline actions. When offline measures are considered, they often focus on attitudinal indicators, for example by mapping online discourse to survey-based measures of climate opinion, rather than concrete behavioral outcomes \citep{loureiro2020sensing, bennett2021mapping}. When such links are attempted, they often rely on coarse proxies such as hashtags, keywords, or community membership \citep{pearce2019social, gaupp2024climate}. As a consequence, we still lack a clear and systematic mapping between online climate discourse and reliable measures of pro-environmental behavior.

% PROPOSAL
To address this gap, we study whether large-scale online climate discourse is associated with differences in offline pro-environmental behavior across European regions. To this end, we combine independent online and offline data sources at the regional level.
On the online side, we use the Climate Change Twitter Dataset \citep{effrosynidis2022climate}, which contains over 15 million climate-related tweets posted between 2006 and 2019. On the offline side, we rely on the 2019 Special Eurobarometer on attitudes towards the environment \citep{ZA7602}, which provides harmonized measures of self-reported pro-environmental actions across European regions.

We aggregate and normalize these two sources at the regional level and examine whether the intensity of online climate discourse is associated with regional levels of offline pro-environmental behavior. 
Our analysis is observational and does not aim to establish causal relationships; rather, it assesses whether online discourse contains a meaningful signal related to collective behavioral patterns.

% SUMMARY OF RESULTS
We find a strong and robust positive association between the density of climate-related tweets and the average number of pro-environmental actions across European regions. This relationship remains stable across multiple specifications and sets of socio-economic controls.

We then extend the analysis using Natural Language Processing tools to distinguish different types of online expressions. While knowledge-related content does not show a clear association with offline behavior, expressions of social support and activism are negatively associated with pro-environmental actions. Although counterintuitive, these results suggest that not all forms of online engagement relate to offline behavior in the same way, and highlight the potential of NLP methods to disentangle different social signals within climate discourse.

Guided by this approach, we address two main research questions. 
First, we ask whether the volume of climate-related discourse on Twitter is associated with differences in pro-environmental behaviors across European regions. 
Second, we investigate whether different social and communicative dimensions of online climate discourse—such as activism, social support, and knowledge exchange—relate differently to offline pro-environmental behaviors. 
Together, these questions aim to assess whether large-scale online discourse contains informative signals about the coordination of pro-environmental behavior beyond what can be captured by traditional socio-economic indicators.

\section{Literature Review / Theoretical Background}

%\edo{Alcune citazioni si ripetono, forse dopo è meglio sfoltire} 

A growing literature exploits social media data to study public attention, opinions, and discourse around climate change and related environmental issues \citep{pearson2016can, pearce2019social, sultana2024systematic}. Twitter, in particular, has been widely used to analyze climate-related discussions, a core component of broader environmental communication, thanks to its high temporal resolution and the availability of geolocation signals, including user-declared profile locations \citep{pearce2019social, gaupp2024climate, fownes2018twitter, kirilenko2014public, kirilenko2015people}. Existing studies have examined tweet volumes, sentiment, and stance to characterize climate change awareness, polarization, and reactions to major events such as extreme weather or social mobilizations \citep{effrosynidis2022climate, effrosynidis2022exploring, crispino2024people, williams2015network, kolic2025chambers, veltri2017climate}. 
These works show that online climate discourse responds to media coverage, socio-economic conditions, and collective events.
They also suggest that Twitter-based indicators can complement traditional survey measures of public concern by providing higher-frequency proxies of attention that align with survey-based measures of climate concern \citep{crispino2024people, loureiro2020sensing, bennett2021mapping}.

Despite these advances, most contributions remain focused on online outcomes alone, such as attention or sentiment, without explicitly linking online discourse to independently measured offline behaviors \citep{pearce2019social, gaupp2024climate, fownes2018twitter}.
When offline measures are considered, they often rely on aggregate attitudinal indicators, for example by mapping online discourse to survey-based measures of climate opinion \citep{loureiro2020sensing, bennett2021mapping}, or on coarse proxies derived from online activity itself \citep{pearce2019social, gaupp2024climate}.
More broadly, digital-trace studies have linked geolocated social media signals to offline outcomes in other domains, including local electricity consumption and conservation-related behaviors \citep{sun2021estimating, crespo2023role}.
As a result, the relationship between large-scale online climate discourse and actual pro-environmental behavior remains insufficiently explored, particularly in a comparative, cross-regional context based on harmonized survey data.

Survey data from the Eurobarometer have also been widely used in quantitative studies of opinions, attitudes, and collective cultural patterns in Europe. Beyond descriptive public-opinion research, Eurobarometer data have been employed to analyze the structure of cultural space, the distribution of opinions across populations, and their implications for social influence and collective behavior \citep{valori2012reconciling, buabeanu2017signs, buabeanu2018ultrametricity}. This makes Eurobarometer a particularly useful source for our purposes, as it provides harmonized cross-national measures that can be linked to broader questions about coordination and behavioral patterns.

From a behavioral perspective, a large body of evidence highlights the importance of social incentives, norms, and network effects in shaping pro-environmental actions. Social influence, network connections, and trust have been shown to play a central role in promoting sustainable behaviors, often complementing or substituting monetary incentives \citep{nguyen2021social}. 
This suggests that pro-environmental behaviors are socially embedded processes that may leave detectable traces in collective communication, making online discourse a potentially informative (though indirect) signal of regional behavioral patterns \citep{chatterjee2024communication}.
In parallel, reviews of climate communication on Twitter emphasize that online discourse serves heterogeneous functions, ranging from informational exchange to mobilization-oriented communication \citep{fownes2018twitter}. 
Building on these perspectives, we examine two social dimensions that plausibly relate to behavioral diffusion and reinforcement, namely \textit{knowledge exchange} and \textit{social support}, and we additionally consider \textit{activism} to capture mobilization-oriented expressions that are prominent in climate-related Twitter discourse \citep{fownes2018twitter}.

To operationalize these dimensions in text, we draw on Natural Language Processing tools that can detect social intent beyond simple volume-based measures. Prior work in computational social science emphasizes that social interactions are multidimensional and cannot be fully captured by frequency alone \citep{deri2018coloring}. Recent NLP frameworks identify distinct social and communicative dimensions in text, including knowledge exchange and social support, and show that these dimensions can relate differently to population-level outcomes \citep{choi2020ten, aiello2022multidimensional}. Complementarily, NLP tools grounded in social movement theory enable the detection of expressions of participation in collective action, providing a more nuanced characterization of activism than keyword-based approaches \citep{pera2025extracting}.

Building on previous literature, this study links large-scale Twitter discourse to survey-based measures of pro-environmental behavior across European regions. By combining aggregate indicators of online climate discourse with NLP-based dimensions capturing activism, social support, and knowledge exchange, the analysis aims to assess whether, and how, online communication patterns are associated with offline pro-environmental actions, while remaining explicitly observational in scope. More broadly, this work contributes to ongoing efforts to integrate large-scale digital traces with social and economic theory to study collective behavior.

\section{Materials and Methods}

\subsection{Data Sources}
To investigate the relationship between online climate discourse and the adoption of pro-environmental behaviors across European regions, we combine two main data sources: the \textit{Climate Change Twitter Dataset} (CCTD) and the \textit{Special Eurobarometer} surveys. Additional socioeconomic and educational indicators are included as confounding factors, sourced from Eurostat and the Eurobarometer itself. All variables are aggregated following the Nomenclature of Territorial Units for Statistics (NUTS) classification.
The NUTS is a hierarchical system for dividing up the economic territory of the EU for the purpose of collection, development, and harmonization of European regional statistics. NUTS1 is the highest level of the NUTS hierarchy and represents major socio-economic regions, or groups of regions, within the EU. The definition of NUTS regions we use throughout this study is based on the 2016 version of the NUTS classification, which includes 27 NUTS regions (EU countries), 92 NUTS1 regions and 281 NUTS2 regions. Our empirical analysis focuses on 185 regions, based on Eurobarometer coverage and coding constraints.

\subsubsection{Twitter Data}
We use the Climate Change Twitter Dataset (CCTD), presented in \cite{effrosynidis2022climate}, which contains 15,789,411 tweets spatially distributed worldwide, spanning 13 years from June 2006 to October 2019. Each tweet is annotated with seven attributes, in addition to the timestamp and unique ID: geolocation (longitude and latitude), climate change stance (believer, neutral, or denier), sentiment, aggressiveness, user gender, deviation from historic temperature, and discussed climate change topic. 
These features were annotated using state-of-the-art machine learning models available at the time of the study (BERT, RNN, LSTM, CNN, SVM, Naive Bayes, VADER, Textblob, Flair, and LDA).
Not all tweets were originally geolocated; however, the authors of the dataset addressed this limitation by extracting location information from users' Twitter profiles. Users often provide their location in a dedicated free-text field, requiring filtering and accurate identification of valid locations. This process successfully geolocated more than a third of the dataset. The dataset authors report that the median haversine distance\footnote{Shortest distance between two points on a sphere, computed from their geographic coordinates.} between coordinates derived from user location metadata and the actual coordinates of a tweet (for those tweets where the coordinates are available) is 24 km. Although this is larger than results found in the relevant literature (13 km in \cite{kirilenko2014public}, 11 km in \cite{kirilenko2015people}), it remains acceptable for our purposes, as it is still smaller than the average NUTS2 region size considered in this work, and comparable with other studies (27.8 km in \cite{sisco2017extreme}).
This dataset has been validated and used in previous studies \citep{effrosynidis2022exploring}.

\subsubsection{Pro-environmental Behaviors Data}
We rely on Eurobarometer (EB), a series of cross-national public opinion surveys conducted on behalf of the European Commission. Running since 1973, EB measures public opinion on a variety of social, political, and economic issues within the European Union.
This study specifically utilizes the 2019 edition of the Special Eurobarometer \citep{ZA7602}, which focuses on the attitudes of European citizens towards the environment. This is the most recent edition of a series focused on the environment, which began in 2007 and was repeated in 2011, 2014, and 2017. 
%All editions contain similar questions, but differences in wording can complicate comparisons.

These surveys include questions on environmental concerns, climate change awareness, and self-reported pro-environmental behaviors, making them particularly useful for examining public engagement with climate and environmental issues at a regional level. The data are statistically representative of the population aged 15 and over in each EU member state, with a sample size of approximately 1,000 respondents per country. Surveys are conducted through face-to-face interviews, using a standardized questionnaire administered by trained interviewers in all EU member states, as well as in candidate and potential candidate countries; however, we focus only on EU member states. The data are publicly available through the GESIS website (\url{www.gesis.org}).

The most relevant survey question for our analysis is: \textit{“In the past 6 months, have you personally done any of the following things to protect the environment?”} Participants could choose from a list of pro-environmental actions, such as \textit{“Recycled waste,” “Avoided using a car,” “Saved energy at home,” “Bought environmentally friendly products,”} etc. The exact list of possible answers is provided in the Appendix.
In 2019, the number of possible actions was 14. 
The dependent variable is the average number of self-reported pro-environmental actions per participant in each region.

For each survey participant, the corresponding NUTS region is recorded. However, the coding of regions is not always consistent with the official NUTS 2016 classification, necessitating manual matching for some regions. Moreover, Germany and Italy, despite being two of the most populous countries in Europe, are only divided into NUTS1 regions, instead of NUTS2 regions, which is the level of detail we target. Therefore, to perform the analysis at a finer scale, we consider all regions at the NUTS2 level, while retaining NUTS1 regions for Germany and Italy. The final dataset covers 185 NUTS2\footnote{Here and throughout, "NUTS2 regions" refers to the combination of true NUTS2 and NUTS1 regions of Germany and Italy.} regions.

Other levels of territorial aggregation include 87 NUTS1 regions and 27 countries (NUTS0), which are used for robustness checks in the Appendix. The total number of NUTS1 regions in the 2016 classification is 92; five regions were not included in the Eurobarometer surveys.

\subsubsection{Confounding Factors}
To assess the effect of possible confounding factors, we consider several socioeconomic variables. To ensure consistency with the main dataset, we extract from the same Eurobarometer survey information on participants' education level and economic situation. Specifically, education is measured by the age at which people stopped full-time studies and socioeconomic status is assessed via self-reported social class. %people->participants, to be consistent with the rest of the text.

Additional variables recorded from the Eurobarometer for further analyses include type of area of residence (rural, suburban, or urban), serving as a proxy for population density; difficulties in paying bills, as a proxy for purchasing power (not directly available in the Eurobarometer dataset); frequency of internet use, as a proxy for digital literacy; and political orientation, which can be used to understand political engagement.

We also use the Eurostat database to compare results with an independent and reliable source. Specifically, we focus on the fraction of the population with tertiary education\footnote{\url{https://ec.europa.eu/eurostat/databrowser/view/TGS00109/default/table}}, that is, the percentage of people who have completed at least a tertiary education program, and on regional purchasing power standards (PPS) per inhabitant\footnote{\url{https://ec.europa.eu/eurostat/databrowser/view/TGS00005/default/table}}, that is, a measure of regional economic output adjusted for price level differences across countries.

%The socioeconomic variables from Eurobarometer and Eurostat differ in definition but are related. We run the analyses with both sets of variables for comparison, however, we primarily focus on the Eurobarometer variables, as they are derived from the same survey as the pro-environmental behaviors data.

\subsection{Data Processing \& Cleaning}

\subsubsection{Twitter Data}
We filter tweets to include only those posted from January 1, 2017 until the end of the dataset (October 2019), and geolocated within the EU, to ensure comparability with the Eurobarometer survey. Tweets were then assigned to corresponding NUTS2 regions according to the NUTS 2016 classification.
Tweets in Germany and Italy were aggregated to the NUTS1 level due to EB survey constraints. The final dataset consists of 331,020 tweets (2.1\% of the original dataset), covering 185 regions.

\begin{figure}[h]
    \centering
    \begin{subfigure}{0.47\textwidth}
        \centering
        \includegraphics[width=\textwidth]{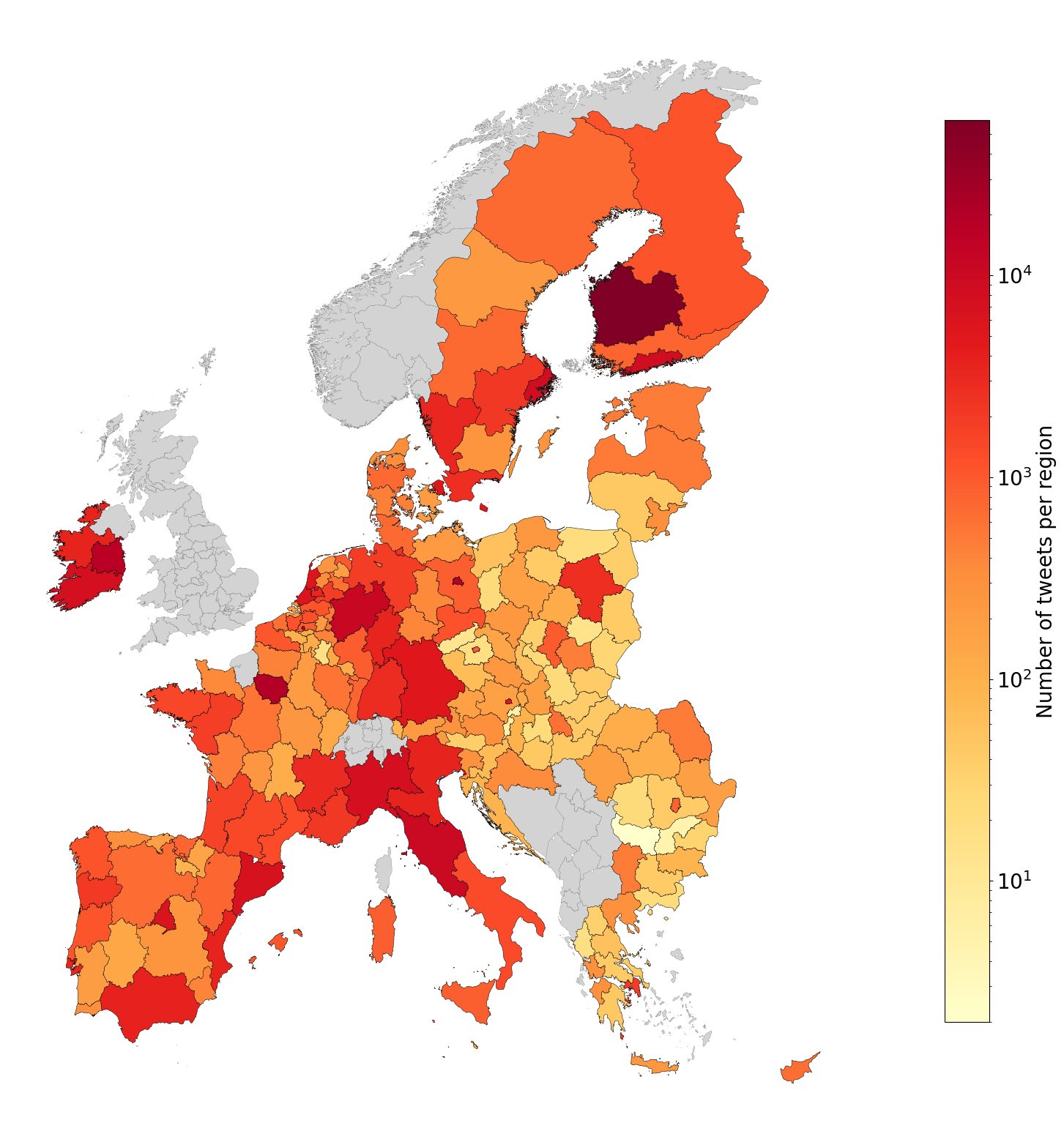}
        \caption{Spatial distribution of the number of tweets.}
        \label{fig:map_tot_tweets_noES70}
    \end{subfigure}
    \hfill
    \begin{subfigure}{0.47\textwidth}
        \centering
        \includegraphics[width=\textwidth]{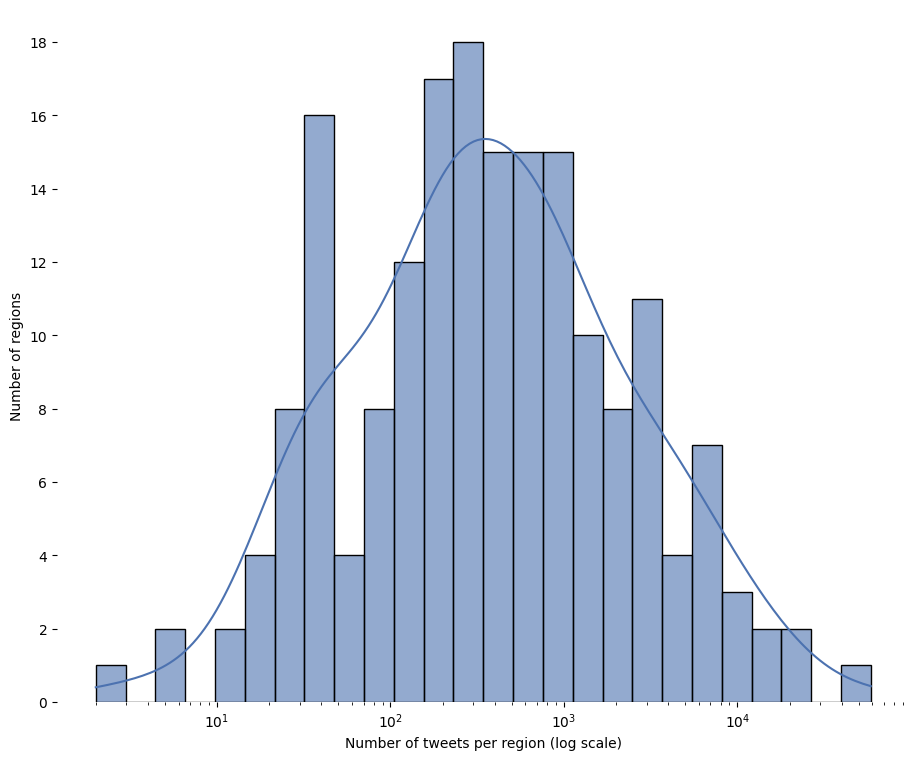}
                \caption{Distribution of the number of tweets per region.}
        \label{fig:dist_tot_tweets_noaxis}
    \end{subfigure}
    \caption{Distribution of the number of tweets in Europe per region. Regions not considered or with no data are shown in grey. Overseas EU territories are not shown.}
    \label{fig:spatial_distribution_combined}
\end{figure}

The spatial distribution of the number of tweets in each region is shown in Figure~\ref{fig:map_tot_tweets_noES70}, while the corresponding distribution across regions is shown in Figure~\ref{fig:dist_tot_tweets_noaxis}.\\

Although each tweet is annotated with a stance (believer, neutral, denier), in this study we use the total number of tweets per region as the main predictor. This captures the overall flow of climate-related information in each region, and leads to more precise regression results than focusing only on believer tweets.

\subsubsection{Pro-environmental behaviors data}

\begin{figure}[h]
    \centering
    \begin{subfigure}{0.47\textwidth}
        \centering
        \includegraphics[width=\textwidth]{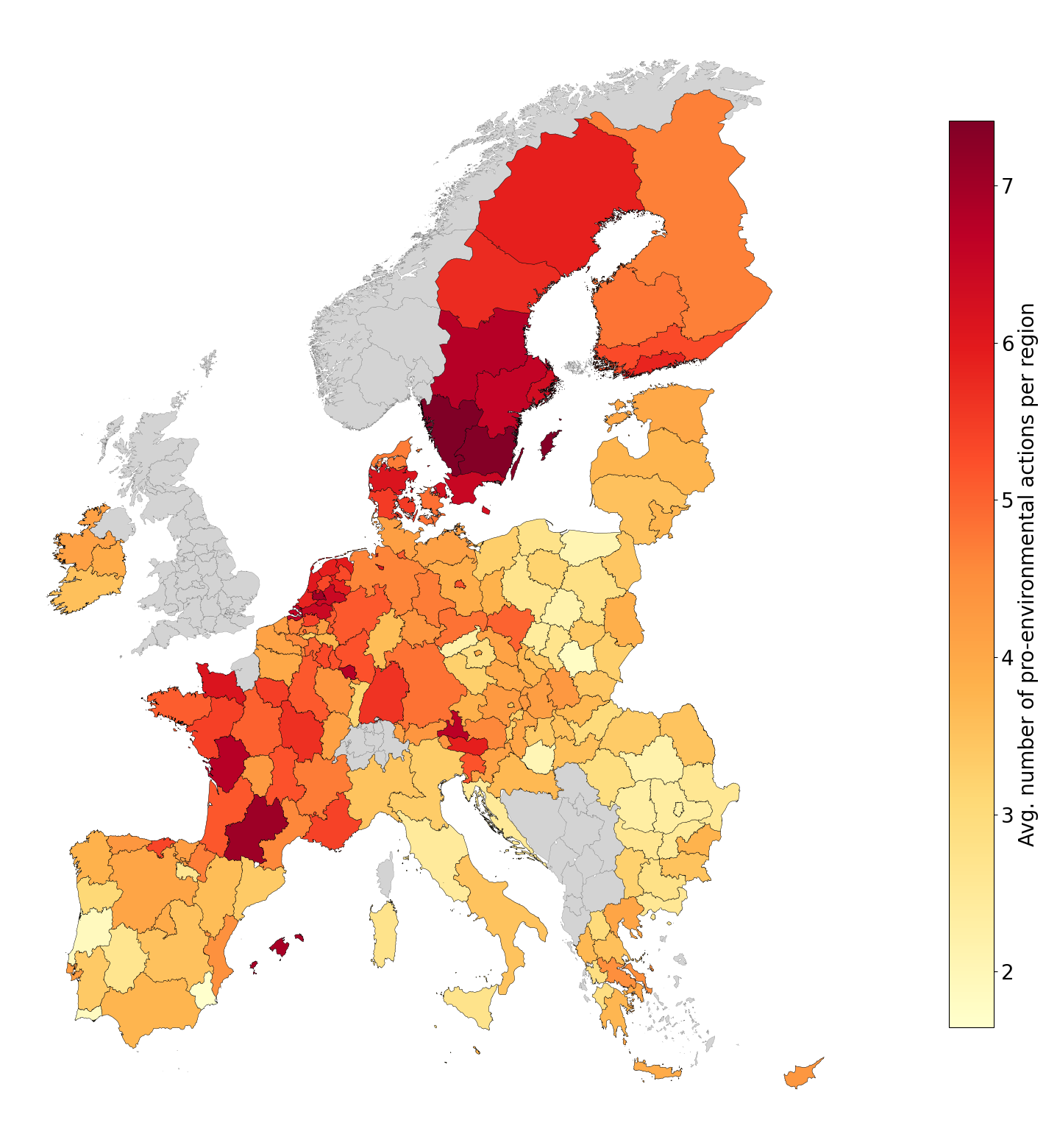}
        \caption{Spatial distribution of the average number of pro-environmental actions, per region.}
        \label{fig:map_avg_proenv}
    \end{subfigure}
    \hfill
    \begin{subfigure}{0.47\textwidth}
        \centering
        \includegraphics[width=\textwidth]{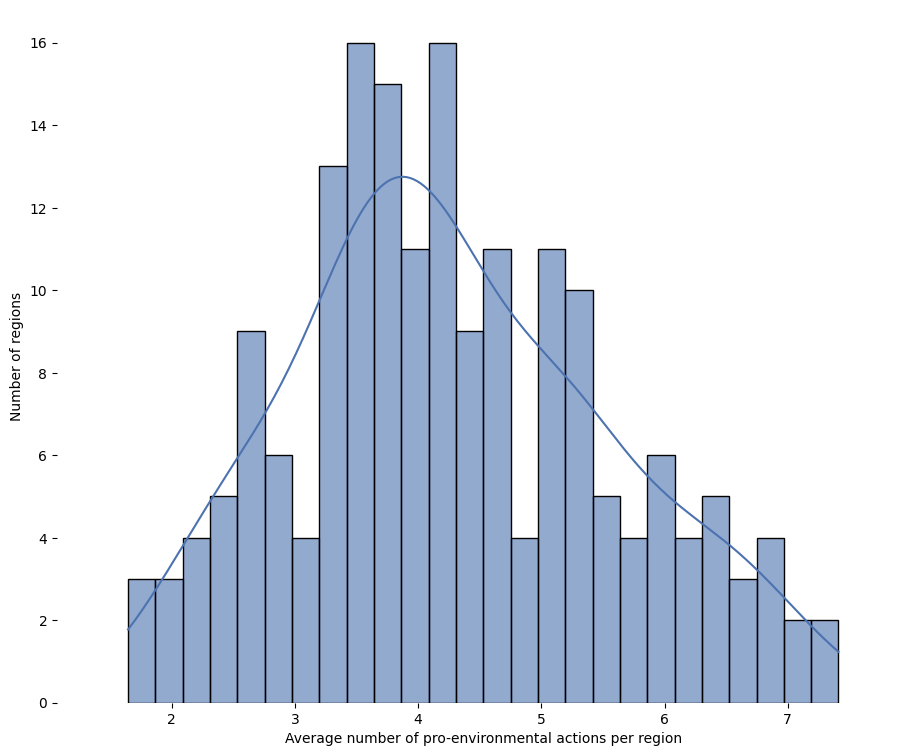}
        \caption{Distribution of the average number of pro-environmental actions, per region.}
        \label{fig:dist_avg_proenv}
    \end{subfigure}
    \caption{Distribution of the average number of pro-environmental actions in 2019 in each region considered in the analysis. Mean=3.253, standard dev.= 0.765. Regions not considered or with no data are shown in grey. Overseas EU territories are not shown, but considered in the analysis.}
    \label{fig:dist_avg_proenv19}
\end{figure}

To obtain the average number of self-reported pro-environmental actions per participant in each region, we use the 2019 Eurobarometer dataset. We compute the sum of the number of actions reported by each participant in each region and then divide it by the number of participants per region. The distribution of the average number of pro-environmental actions per participant in each region is shown in Figure \ref{fig:dist_avg_proenv19}.

\subsubsection{Confounding factors}
A similar procedure is applied to the confounding factors obtained from the Eurobarometer, for which we compute the average of each variable across respondents in each region. Eurostat variables, instead, are already available as regional averages.

\subsection{Data Normalization}
\subsubsection{Twitter Data}
To use the count of tweets as a predictor, it must be properly normalized.
A natural approach is to divide the number of climate-related tweets by the total number of tweets posted in that region. Unfortunately, the information on the total volume of tweets per region is not publicly available and, due to the new Twitter/X policy, it is no longer accessible.
We therefore consider the density of tweets in each region, defined as the average number of tweets per capita obtained by normalizing counts by the regional population. This practice allows for a comparison between regions with different population sizes. This normalization procedure has been commonly applied in spatial studies of social media:~\cite{longley2015geotemporal} computed the number of Twitter users per capita across London census areas to account for population differences, while \cite{malik2015population} analyzed the ratio of geotag users to population across U.S. census block groups. Both approaches rely on tweet or user densities, comparable to our measure of tweets per capita. Moreover, this procedure is coherent with the normalization of the pro-environmental behaviors data, which already represent the average per capita number of actions in each region.

The distribution of the fraction of tweets w.r.t. the population in each NUTS2 region is shown in \SFig{fig:dist_perc_nuts1}. 
\SFig{fig:dist_perc_nuts1_log} shows that the distribution of the fraction of tweets w.r.t. the population in each NUTS2 region is approximately log-normal, which means that the distribution of the logarithm of the fraction of tweets w.r.t. the population in each NUTS2 region is approximately normal and more suitable for our computations. 
This is expected, since the number of tweets in a region can grow without a fixed upper bound. Some regions have a very large number of tweets, while many others have a small number of tweets. 
Therefore, here and with the other quantities that can grow indefinitely, we use the logarithm.
Our main predictor is the logarithm of the number of tweets per region, normalized by the regional population.

Finally, we standardize the variables by calculating their z-score, a measure of how many standard deviations a data point is from the mean. We use standardized scores in the correlation and regression analyses.

\subsubsection{Pro-environmental behaviors data}
The average number of pro-environmental actions per participant in each region is already normalized by the number of participants. It represents the average number of actions per capita in each region, coherently with the normalization of the number of tweets.
Therefore, it is used as the main dependent variable for the adoption of pro-environmental behaviors.
In this case, since the average number of pro-environmental actions per participant is bounded between 0 and 14, it cannot grow indefinitely, and therefore it is not necessary to use the logarithm.

\begin{figure}[h]
    \centering
    \begin{subfigure}{0.47\textwidth}
        \centering
        \includegraphics[height=1.1\textwidth]{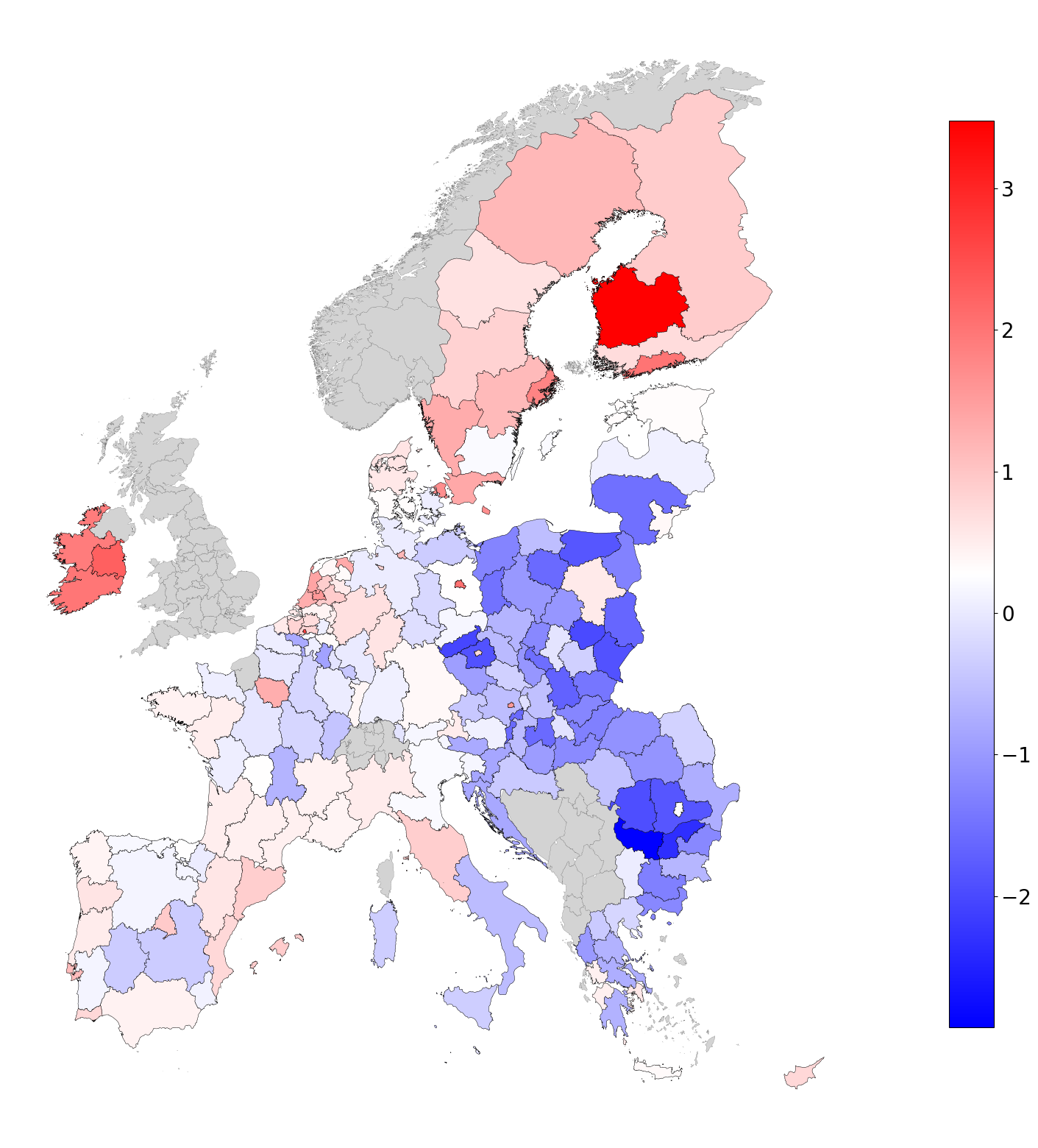}
        \caption{Fraction of total tweets per region (log)}
        \label{fig:map_tot_tweets}
    \end{subfigure}
    \begin{subfigure}{0.47\textwidth}
        \centering
        \includegraphics[height=1.1\textwidth]{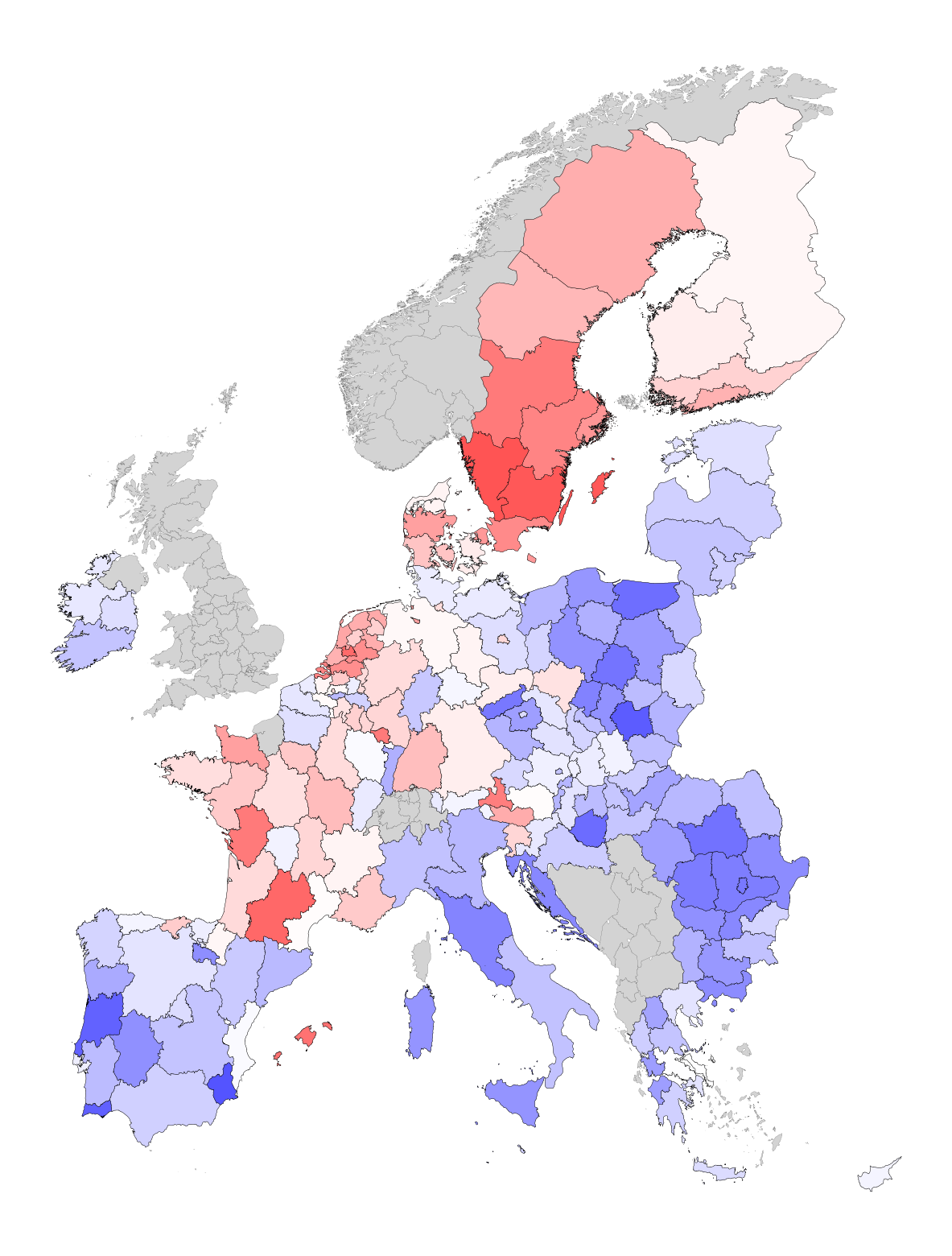}
        \caption{Average number of pro-environmental actions}
        \label{fig:map_actions}
    \end{subfigure}
    \caption{Spatial distribution of the z-score of the fraction of climate-related tweets per region (log) (a) and the average number of pro-environmental actions (b) in each NUTS2 region. The regions not considered or with no data are shown in grey. Overseas EU territories are not shown, but are considered in the analysis.}
    \label{fig:dist_avg}
\end{figure}

In order to standardize the variable, also in this case, we use the z-score. The spatial distributions of the two standardized variables are shown in Figure \ref{fig:dist_avg}.

\subsubsection{Confounding factors}

Regarding the confounding factors computed through the Eurobarometer, as for the pro-environmental behaviors data, all variables are already averages and are therefore used as they are in the analysis. However, the level of education is not bounded and needs to be logarithmically transformed, while all other variables are bounded.
%The social class, instead, is bounded between 1 and 5, and therefore it is not necessary to use the logarithm. 
%The distribution of the social class in each NUTS2 region is shown in Figure \ref{fig:dist_social_class}.
%The distribution of the difficulties in paying the bills has a tail on the left and necessitates a logarithmic transformation of its complement. \edo{I will explain this better later.}
%For all the variables, we computed the z-score in order to standardize them to perform the regression analysis.

Regarding the confounding factors computed through the Eurostat database, the regional gross domestic product (PPS per inhabitant) is not bounded and can grow indefinitely, and we therefore apply the logarithm. %The normal shape of the distribution of the logarithm justifies this choice, as shown in Figure \ref{fig:dist_ESmain} Left. 
The fraction of the population with tertiary education, instead, is bounded between 0 and 1. The distributions of the two final variables in Figure~\ref{fig:dist_ESmain} are approximately normal.

For all confounding factors, as for the other variables, we compute the z-score in order to standardize them for the regression analysis.

\subsection{Model}
\subsubsection{Variables}

For the core analysis, we focus on the number of tweets and the average number of pro-environmental actions per participant, aggregated at the NUTS2 level, with the exception of Germany and Italy, where we use NUTS1. For each variable, we compute the z-score.

The main variables are summarized in Table~\ref{tab:variables_summary}.

\begin{table}[h]
    \centering
    % Increase line spacing for this table only
    \renewcommand{\arraystretch}{1.4}
    \begin{tabular}{p{0.23\textwidth} p{0.1\textwidth} p{0.65\textwidth}}
        \toprule
        \textbf{Category} & \textbf{Variable} & \textbf{Description}\\
        \midrule
        Independent variable & $DenTweets$ & Logarithm of the number of tweets between 2017 and 2019 in each region, normalized by the regional population in 2019.\\
        \addlinespace
        \midrule
        Dependent variable & $ProEnv_{19}$ & Average number of self-reported pro-environmental actions (from Eurobarometer 2019) per capita, for each European region.\\
        \addlinespace
        \midrule
        Main confounding \hspace{0.3cm} factors (Eurobarometer) & $EB_{EDU}$ & Logarithm of the average number of years of full-time education.\\
        & $EB_{SOC}$ & Average social class of Eurobarometer respondents.\\
        \addlinespace
        \midrule
        Additional confounding factors (Eurobarometer) & $EB_{ECON}$ & Fraction of Eurobarometer respondents reporting difficulties in paying bills.\\
        & $EB_{CITY}$ & Type of area of residence (rural, suburban, or urban).\\
        & $EB_{INT}$ & Frequency of internet use.\\
        & $EB_{POL}$ & Political orientation of Eurobarometer respondents.\\
        \addlinespace
        \midrule
        Alternative confounding factors (Eurostat) & $ES_{EDU}$ & Fraction of the population with tertiary education.\\
        & $ES_{PPS}$ & Logarithm of the standardized regional gross domestic product (PPS per inhabitant).\\
        \bottomrule
    \end{tabular}
    \renewcommand{\arraystretch}{1}
    \caption{Summary of the main variables used in the analysis.}
    \label{tab:variables_summary}
\end{table}

\subsubsection{Statistical Analysis}

We begin by measuring the association between dependent and independent variables using Pearson’s correlation coefficient, assessing statistical significance through the corresponding t-test and p-value.

To assess the robustness of the observed correlations and control for potential confounding factors, we perform a series of Ordinary Least Squares (OLS) regressions. These models include, alongside the main Twitter-based predictors, well-established socio-economic variables, explained above.
\begin{equation}
ProEnv_{19,r} = \alpha + \beta \, DenTweets_r + \gamma^\top X_r + \varepsilon_r
\end{equation}
where $r$ denotes the region, $\alpha$ is the intercept, $\beta$ is the coefficient associated with tweet density, $X_r$ represents the vector of control variables, $\gamma$ is the corresponding vector of coefficients, and $\varepsilon_r$ is the error term. Depending on the specification, $X_r$ includes the main Eurobarometer controls, the additional Eurobarometer controls, or the alternative Eurostat controls. Throughout the Results section, for readability, we refer to coefficients using the names of the corresponding variables.

This allows us to estimate the net contribution of the Twitter signal relative to these background variables by comparing the magnitude and significance of the regression coefficients. 
We first consider the independent variable together with the main confounding factors from Eurobarometer, and we estimate regressions for every possible combination of the predictors.

We conduct several additional tests. These include adding further control variables from Eurobarometer, replacing Eurobarometer controls with alternative Eurostat indicators.
These additional tests are intended to assess whether the observed association depends on a specific set of controls, or on particular operational choices in the construction of the online signal.

Furthermore, we test the consistency of our findings across different scales of territorial aggregations (NUTS0, NUTS1).
We also distinguish between two behavioral dimensions: the intensive component, capturing the average number of actions among already-engaged individuals (i.e., those who reported at least one action), and the extensive component, defined as the proportion of active participants in each region.

We also repeat the analysis aggregating the tweets to their respective Twitter account and computing the number of users instead of the number of tweets. 
We also assess the sensitivity of the results to the choice of the tweet collection window by varying the starting date of the dataset.

Finally, we test the robustness of the results by removing outliers, that is, regions with the lowest number of tweets and the fewest Eurobarometer participants at the NUTS2 level, and by repeating the analysis considering only tweets classified as expressing a believer stance towards climate change.

\subsection{NLP Analysis}

To move beyond aggregate measures of tweet volume, our analysis requires indicators that distinguish different forms of online communication that may plausibly relate to collective behavioral coordination.
In particular, we focus on dimensions capturing activism, informational exchange and social support, which are often discussed as relevant mechanisms in the diffusion and reinforcement of prosocial behaviors, as in \citet{nguyen2021social}.
To complement our analysis, we use Natural Language Processing (NLP) tools designed to capture the social and mobilization intent expressed in tweets. 
We focus on three distinct dimensions identified in the literature as relevant to collective behavioral dynamics: \textit{activism}, \textit{knowledge}, and \textit{support}. These dimensions capture, respectively, mobilization-oriented expressions of collective action, informational exchange, and expressions of emotional or practical support.

We use a tool developed by \citet{pera2025extracting} to detect expressions of participation in collective action in social media texts. Grounded in the theoretical framework of social movement mobilization, it identifies four incremental levels of participation: recognizing collective issues, engaging in calls-to-action, expressing intention of action, and reporting active involvement. Trained on a crowdsourced and augmented dataset of Reddit comments, the model is topic-agnostic and recognizes any expression of collective participation, regardless of the underlying cause. It therefore provides a robust measure of users’ engagement in collective discourse.

To operationalize the other two dimensions in text, we rely on the framework introduced by \citet{deri2018coloring} and \citet{choi2020ten}, which proposes and validates ten fundamental social dimensions of human communication: \textit{knowledge exchange}, \textit{social support}, \textit{trust}, \textit{power}, \textit{status}, \textit{similarity}, \textit{identity}, \textit{romance}, \textit{fun}, and \textit{conflict}. Consistently with this objective, we restrict our analysis to the \textit{knowledge} and \textit{support} dimensions.

For each dimension, the corresponding classifier assigns a continuous score between 0 and 1, estimating the presence of that social intent in the text. To construct region-level predictors, we adopt a quantile-based thresholding approach, defining thresholds relative to the empirical distribution of scores across the entire dataset rather than using a fixed cut-off. This strategy preserves a constant fraction of selected tweets for each dimension while maintaining cross-regional comparability. We then examine how the correlation between the regional prevalence of each dimension and the average number of pro-environmental actions varies across a wide range of quantile thresholds. Based on this inspection, we select thresholds that yield stable and interpretable associations for \textit{activism} and \textit{support}. For \textit{knowledge exchange}, no threshold produces a meaningful or stable association, and we therefore do not construct a regional predictor for this dimension.

We then compute, for each NUTS2 region, the number of tweets classified above the corresponding threshold for each dimension, normalized by the total number of tweets in that region. These normalized quantities represent the regional prevalence of each social intent. Finally, they are correlated with the average number of pro-environmental actions and used as independent variables in multivariate OLS regressions with the same confounding factors described above.

\section{Results}

\subsection{Correlation Analysis}
\begin{figure}[h]
        \centering
        \includegraphics[width=1\textwidth]{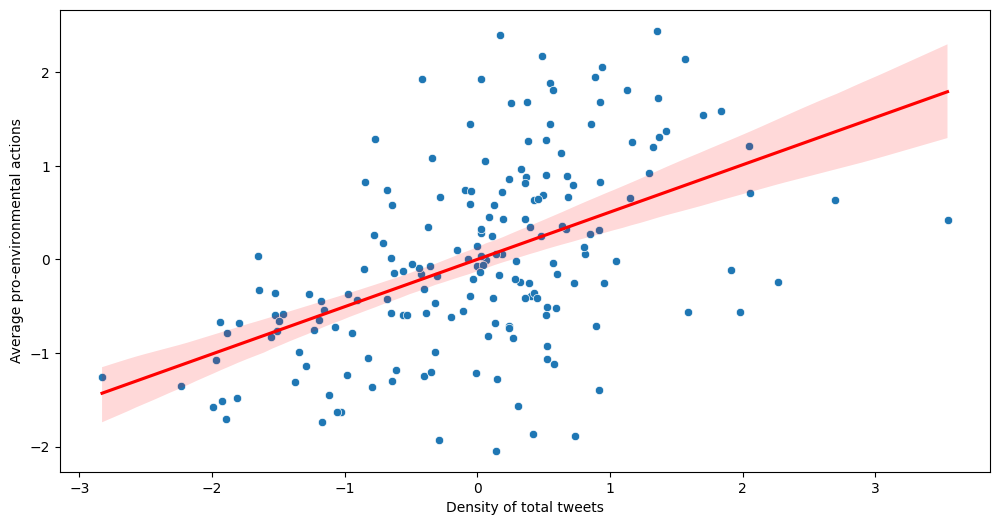}
        \caption{Scatterplot of $DenTweets$, logarithm of the density of tweets per region normalized by the regional population, and $ProEnv_{19}$, average number of self-reported pro-environmental actions (from Eurobarometer 2019) per capita, for each European region. Z-scores are computed for both variables. Pearson's correlation coefficient is $r=0.506$, with a p-value of $p=2.079\cdot 10^{-13}$. 185 regions considered.}
        \label{fig:corr_tot}
\end{figure}

We first examine the association between the logarithm of tweet density per region, normalized by the regional population, and the average number of pro-environmental actions per participant in 2019. For both variables, we compute the z-score. As shown in Figure~\ref{fig:corr_tot}, the two variables are positively associated: regions with higher levels of climate-related Twitter activity tend to report higher average levels of pro-environmental behavior. The Pearson correlation coefficient is $r=0.506$, with a p-value of $p=2.079\cdot 10^{-13}$, indicating a strong and highly significant relationship. Although this correlation does not account for potential confounding factors, it provides a clear first indication that tweet density contains a meaningful signal related to regional differences in offline pro-environmental action.

\subsection{Regression with main confounding factors from Eurobarometer}
To assess whether the observed association persists when accounting for key socio-economic characteristics, we perform an Ordinary Least Squares (OLS) regression analysis. We use as independent variables the logarithm of tweet density per region, $DenTweets$, the logarithm of the average years of full-time education, $EB_{EDU}$, and the social class of Eurobarometer respondents, $EB_{SOC}$. The distributions of these standardized variables are shown in \SFig{fig:dist_perc_nuts1} and \ref{fig:dist_tweet_proenv}. %and the correlations between all the standardized independent variables 
The coefficients of the regressions are reported in Table~\ref{tab:regression_mainEB}.

The results show that tweet density remains strongly associated with pro-environmental actions when the main Eurobarometer controls are included. In the single-predictor models, the standardized coefficient of $DenTweets$ is $0.506$, compared with $0.445$ for $EB_{EDU}$ and $0.495$ for $EB_{SOC}$. When tweet density is combined with each control separately, its coefficient stays positive, highly significant, and larger than that of the other variable. The model with tweet density alone also explains a substantial share of the variation in pro-environmental behavior ($R^2=0.256$), close to the model that includes education and social class together ($R^2=0.311$). In the full model with all three predictors, the coefficient of $DenTweets$ remains statistically significant and larger than those of both $EB_{EDU}$ and $EB_{SOC}$, and the $R^2$ rises to 0.422. Overall, these results suggest that tweet density captures part of the regional variation in pro-environmental behavior beyond what is explained by standard socio-economic controls.

\begin{table}[h]
    \begin{tabular}{lccccccc}
        \toprule
        \textbf{$ProEnv_{19}$} & (1) & (2) & (3) & (4) & (5) & (6) & (7)\\
        \midrule
        \textbf{$DenTweets$} & $0.506^{***}$ &  &  & $0.406^{***}$ & $0.398^{***}$ &  & $0.356^{***}$\\
         & (0.064) &  &  & (0.063) & (0.060) &  & (0.060)\\

        \textbf{$EB_{EDU}$} &  & $0.445^{***}$ &  & $0.317^{***}$ &  & $0.284^{***}$ & $0.199^{***}$\\
         &  & (0.066) &  & (0.063) &  & (0.068) & (0.064)\\

        \textbf{$EB_{SOC}$} &  &  & $0.495^{***}$ &  & $0.383^{***}$ & $0.372^{***}$ & $0.309^{***}$\\
         &  &  & (0.064) &  & (0.060) & (0.068) & (0.064)\\

        \midrule
        Number of regions & 185 & 185 & 185 & 185 & 185 & 185 & 185\\
        $R^2$ & 0.256 & 0.198 & 0.245 & 0.346 & 0.391 & 0.311 & 0.422\\
        \bottomrule
    \end{tabular}

    \caption{Regression results for the average number of pro-environmental actions ($ProEnv_{19}$). The main predictor is the number of tweets normalized by the regional population ($DenTweets$), while confounding factors include the average number of years of full-time education ($EB_{EDU}$) and the average social class ($EB_{SOC}$), both computed from the Eurobarometer. For all variables, the z-score is considered. Reported coefficients are shown with standard errors in parentheses. Regions considered at NUTS2 level. * p $<$ 0.1, ** p $<$ 0.05, *** p $<$ 0.01. }
    \label{tab:regression_mainEB}
\end{table}

\subsection{Intensive vs. Extensive contribution}

We next examine more closely the positive association between tweet density and pro-environmental behavior by decomposing pro-environmental behavior into two complementary dimensions: an \textit{intensive} component, capturing the average number of actions among already active individuals, and an \textit{extensive} component, defined as the share of active respondents in each region.

Using the standard definition of activity, that is, at least one reported action, the extensive component shows limited variation across regions, since participation rates are close to saturation in several cases (\SFig{fig:int_ext_ge1}). To obtain a more informative decomposition, we therefore adopt a stricter definition of activity and classify respondents as active only if they reported at least three actions. This threshold reduces ceiling effects while preserving sufficient regional variation.

\begin{figure}[h]
    \centering
    \begin{subfigure}[b]{0.48\linewidth}
    \includegraphics[width=\linewidth]{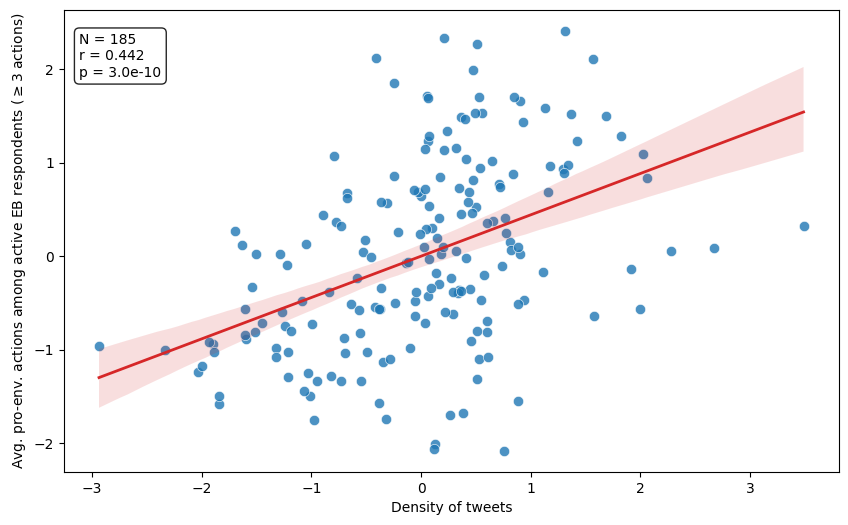}
    \caption{Intensive: average number of actions among respondents with $\geq3$ actions.}
    \label{fig:int_ge3}
    \end{subfigure}
    \hfill
    \begin{subfigure}[b]{0.48\linewidth}
    \includegraphics[width=\linewidth]{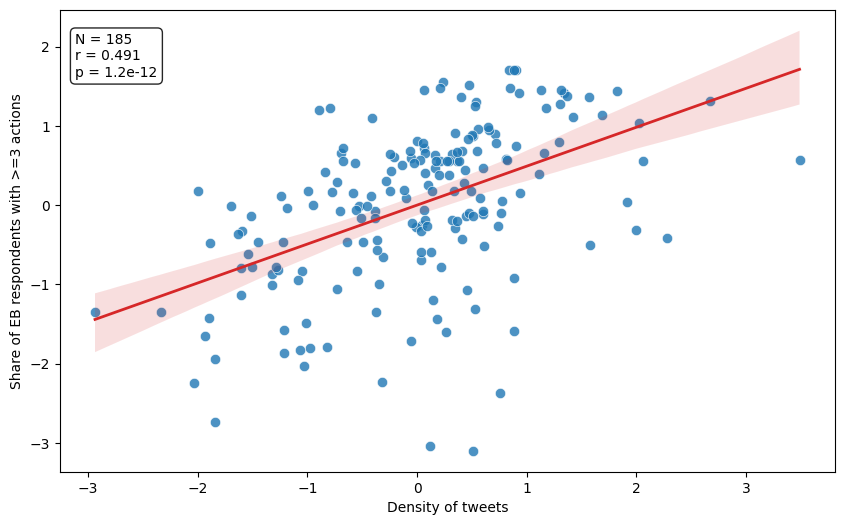}
    \caption{Extensive: share of respondents with at least three \\ actions.}
    \label{fig:ext_ge3}
    \end{subfigure}
    \caption{Correlation between tweet density and pro-environmental behavior for active respondents (threshold $\geq$3). The positive and significant correlations for both the intensive ($r=0.44$) and extensive ($r=0.49$) components confirm that the overall effect combines wider engagement with stronger individual commitment.}
    \label{fig:int_ext_ge3}
\end{figure}

Figure~\ref{fig:int_ext_ge3} shows that, under this definition, both the intensive and extensive components are positively and significantly correlated with tweet density. This suggests that the overall association reflects both a broader diffusion of pro-environmental behaviors and higher levels of engagement among already active individuals. For completeness, we report the corresponding results obtained with the original threshold of one action in \SFig{fig:int_ext_ge1}, which yield qualitatively similar positive associations.

Overall, this decomposition suggests that regions with higher levels of online climate discourse are characterized not only by a larger share of individuals engaging in pro-environmental actions, but also by greater behavioral commitment among those who do so.

The analyses above establish the core relationship between online climate discourse and offline pro-environmental behavior. We now turn to a more fine-grained analysis of the types of online expressions associated with this relationship.

\subsection{Regression analysis on NLP outputs}

We complement the aggregate volume analysis with NLP-based indicators that capture distinct social and communicative dimensions of tweets, to investigate whether different types of online climate discourse are differentially associated with offline pro-environmental behavior.

For each tweet and for each dimension, the NLP models return a continuous score between 0 (no presence of the dimension) and 1 (strong presence). To construct region-level predictors, we focus on the relative prevalence of each dimension by computing, for each region, the share of tweets whose score exceeds a given threshold, normalized by the total number of climate-related tweets in that region.

For each dimension, we report the results obtained using the quantile-based threshold selection procedure described in the Methods section.

The results are presented separately for each dimension, while the score distributions are reported in \SFig{fig:dim_dist}. For activism and social support we report the behavior of correlations across quantiles, \SFig{fig:corr_plot_act} and \SFig{fig:corr_plot_supp} respectively, and the corresponding regression results. For knowledge exchange, we show that no threshold produces a meaningful or stable association, and therefore no regression-based predictor is constructed for this dimension.

\subsubsection{Activism}

The distribution of activism scores shown in \SFig{fig:act_dist} follows a clear bimodal shape, with a large peak at 0 and a smaller one at 1.
To select an appropriate threshold, we computed the Pearson correlation between the regional fraction of activist tweets (above a given quantile) and the average number of pro-environmental actions, varying the cutoff between the 50th and the 99.9th percentile (\SFig{fig:corr_plot_act}).

The quantile that maximizes the absolute value of the correlation ($r = -0.227$, $p = 0.0026$) is 0.95, as shown in \SFig{fig:corr_plot_act}, and it remains significant when using Eurobarometer controls (Table~\ref{tab:regression_act_EB95}). This indicates a stable negative association between the prevalence of activist tweets and the average number of pro-environmental actions, even after accounting for the main socio-economic controls. In the robustness section, we also show that the negative association is not specific to this cutoff by repeating the analysis at a lower threshold (0.80), which yields qualitatively consistent results across specifications.

\begin{table}[h]
    \begin{tabular}{lccccccc}
        \toprule
        \textbf{$ProEnv_{19}$} & (1) & (2) & (3) & (4) & (5) & (6) & (7)\\
        \midrule
        \textbf{$ActTweets$} & $-0.227^{***}$ &  &  & $-0.203^{***}$ & $-0.119^{*}$ &  & $-0.133^{**}$\\
         & (0.074) &  &  & (0.067) & (0.068) &  & (0.065)\\

        \textbf{$EB_{EDU}$} &  & $0.442^{***}$ &  & $0.430^{***}$ &  & $0.280^{***}$ & $0.288^{***}$\\
         &  & (0.069) &  & (0.067) &  & (0.071) & (0.071)\\

        \textbf{$EB_{SOC}$} &  &  & $0.490^{***}$ &  & $0.462^{***}$ & $0.367^{***}$ & $0.333^{***}$\\
         &  &  & (0.067) &  & (0.068) & (0.071) & (0.073)\\

        \midrule
        Number of regions & 173 & 173 & 173 & 173 & 173 & 173 & 173\\
        $R^2$ & 0.052 & 0.195 & 0.240 & 0.236 & 0.254 & 0.304 & 0.321\\
        \bottomrule
    \end{tabular}

    \caption{Regression results for the average number of pro-environmental actions ($ProEnv_{19}$). The main predictor is the number of activist tweets (above the 95th percentile) normalized by the number of CCTD tweets in the region ($ActTweets$), while confounding factors include the average social class declared by the EB participants in that region and the average regional education level in 2019, both computed from the Eurobarometer. For all variables, we computed the z-score. Reported coefficients are shown with standard errors in parentheses. Regions considered at NUTS2 level. * p $<$ 0.1, ** p $<$ 0.05, *** p $<$ 0.01. }
    \label{tab:regression_act_EB95}
\end{table}

\subsubsection{Social Support}

The distribution of social support scores is shown in \SFig{fig:supp_dist} and is heavily skewed toward zero (note the logarithmic scale on the y-axis). To select an appropriate threshold, we computed the Pearson correlation between the regional share of support tweets (above a given quantile) and the average number of pro-environmental actions, varying the cutoff between the 50th and the 99.9th percentile (\SFig{fig:corr_plot_supp}).

The correlation curve shows a clear and stable minimum at the 0.97 quantile, with the strongest negative correlation ($r = -0.256$) and the lowest p-values (0.00134). It remains significant when using Eurobarometer controls (Table~\ref{tab:regression_supp_EB}). Compared with activism, the negative pattern for social support is even clearer and more stable across thresholds. At this cutoff, the support signal is consistently negative and significant, indicating that regions with higher levels of individual pro-environmental actions tend to show fewer expressions of emotional or practical support in their online conversations.

\begin{table}[h]
    \begin{tabular}{lccccccc}
        \toprule
        \textbf{$ProEnv_{19}$} & (1) & (2) & (3) & (4) & (5) & (6) & (7)\\
        \midrule
        \textbf{$SuppTweets$} & $-0.256^{***}$ &  &  & $-0.242^{***}$ & $-0.233^{***}$ &  & $-0.230^{***}$\\
         & (0.078) &  &  & (0.069) & (0.068) &  & (0.065)\\

        \textbf{$EB_{EDU}$} &  & $0.463^{***}$ &  & $0.455^{***}$ &  & $0.300^{***}$ & $0.297^{***}$\\
         &  & (0.072) &  & (0.069) &  & (0.075) & (0.073)\\

        \textbf{$EB_{SOC}$} &  &  & $0.495^{***}$ &  & $0.484^{***}$ & $0.358^{***}$ & $0.349^{***}$\\
         &  &  & (0.070) &  & (0.068) & (0.075) & (0.073)\\

        \midrule
        Number of regions & 154 & 154 & 154 & 154 & 154 & 154 & 154\\
        $R^2$ & 0.066 & 0.214 & 0.245 & 0.273 & 0.299 & 0.316 & 0.369\\
        \bottomrule
    \end{tabular}

    \caption{Regression results for the average number of pro-environmental actions ($ProEnv_{19}$). The main predictor is the number of support tweets (above the 97th percentile) normalized by the number of CCTD tweets in the region ($SuppTweets$), while confounding factors include the average social class declared by the EB participants in that region and the average regional education level in 2019, both computed from the Eurobarometer. For all variables, we computed the z-score. Reported coefficients are shown with standard errors in parentheses. Regions considered at NUTS2 level. * p $<$ 0.1, ** p $<$ 0.05, *** p $<$ 0.01. }
    \label{tab:regression_supp_EB}
\end{table}

\subsubsection{Knowledge Exchange}

The distribution of knowledge scores is shown in \SFig{fig:know_dist} and is strongly skewed toward high values, with many tweets scoring above 0.7, consistent with the fact that climate-related discussions often contain informational content.

As anticipated in the Methods section, no quantile threshold yields a meaningful or stable association between the regional share of knowledge-related tweets and the average number of pro-environmental actions (\SFig{fig:corr_plot_know}). Correlations remain close to zero across the entire range of tested thresholds, and the corresponding $p$-values are consistently far from standard significance levels.

For this reason, we do not construct a regional predictor for the knowledge dimension and do not report regression results for this case. At least in our dataset, knowledge-related content does not appear to explain cross-regional differences in offline pro-environmental actions.

\section{Robustness Checks}

\subsection{Regression with different confounding factors}
\subsubsection{Additional confounding factors from Eurobarometer}

We further assess the robustness of the main results by augmenting the baseline specification with additional individual-level covariates from the Eurobarometer, capturing economic conditions, urbanization, digital engagement, and political orientation.

When considered individually, fewer reported difficulties in paying bills and higher frequency of internet use are positively associated with pro-environmental actions, while a more right-leaning political orientation is negatively associated. The type of area of residence does not display a clear relationship with pro-environmental behavior.

Crucially, when all additional covariates are included jointly with tweet density and the main controls for education and social class, the core result remains unchanged. The density of tweets continues to be positively and strongly associated with pro-environmental actions, with a coefficient that is comparable in magnitude to, and in some cases larger than, those of the socio-economic controls. This confirms that the observed association is not driven by omitted individual-level characteristics captured in the survey.

Detailed regression results are reported in \STab{tab:regression_addEB}.

\subsubsection{Alternative confounding factors from Eurostat}

To verify the robustness of the results with an independent data source, we replaced the Eurobarometer controls with Eurostat indicators of regional education and purchasing power. When considered alone, both the fraction of population with tertiary education ($ES_{EDU}$) and the regional purchasing power per capita ($ES_{PPS}$) are strongly and positively associated with the number of pro-environmental actions.

When adding the density of tweets ($DenTweets$) to the model, its coefficient remains positive, highly significant, and stable across specifications. In the joint regression including all controls, a one–standard-deviation increase in tweet density is associated with a 0.25 SD increase in pro-environmental actions ($p < 0.05$), while both Eurostat variables remain significant and positively correlated. The full model explains around 30\% of the variance ($R^2 = 0.306$).

Because all variables are standardized, the coefficients are directly comparable, showing that the association between the density of tweets and pro-environmental behavior is robust and slightly stronger than that of key socio-economic predictors such as education and purchasing power. This confirms that the observed relationship is not driven by regional wealth or educational differences (the closest equivalents to the main confounding factors from Eurobarometer available in Eurostat) but rather reflects an independent and stable signal.

\begin{table}[h]
    \begin{tabular}{lccccccc}
        \toprule
        \textbf{$ProEnv_{19}$} & (1) & (2) & (3) & (4) & (5) & (6) & (7)\\
        \midrule
        \textbf{$DenTweets$} & $0.506^{***}$ &  &  & $0.351^{***}$ & $0.304^{***}$ &  & $0.247^{**}$\\
         & (0.064) &  &  & (0.083) & (0.090) &  & (0.095)\\

        \textbf{$ES_{EDU}$} &  & $0.467^{***}$ &  & $0.235^{***}$ &  & $0.238^{***}$ & $0.160^{*}$\\
         &  & (0.065) &  & (0.083) &  & (0.085) & (0.089)\\

        \textbf{$ES_{PPS}$} &  &  & $0.499^{***}$ &  & $0.280^{***}$ & $0.338^{***}$ & $0.213^{**}$\\
         &  &  & (0.064) &  & (0.090) & (0.085) & (0.097)\\

        \midrule
        Number of regions & 185 & 185 & 185 & 185 & 185 & 185 & 185\\
        $R^2$ & 0.256 & 0.218 & 0.249 & 0.287 & 0.293 & 0.280 & 0.306\\
        \bottomrule
    \end{tabular}

    \caption{Regression results for the average number of pro-environmental actions ($ProEnv_{19}$). The main predictor is the number of tweets normalized by the regional population ($DenTweets$), while confounding factors include the fraction of regional population with tertiary education in 2019 ($ES_{EDU}$) and regional purchasing power per capita ($ES_{PPS}$), both computed from Eurostat. All variables are standardized; logs are applied where appropriate. Reported coefficients are shown with standard errors in parentheses. Regions considered at NUTS2 level. * p $<$ 0.1, ** p $<$ 0.05, *** p $<$ 0.01. }
    \label{tab:regression_tweets_mainES}
\end{table}

% New, less technical and shorter text for this subsection, as per instructions:
\subsection{Different scales of territorial aggregation: NUTS1 and NUTS0}

\begin{figure}[h]
    \centering
    \begin{subfigure}{0.49\linewidth}
        \centering
        \includegraphics[width=\linewidth]{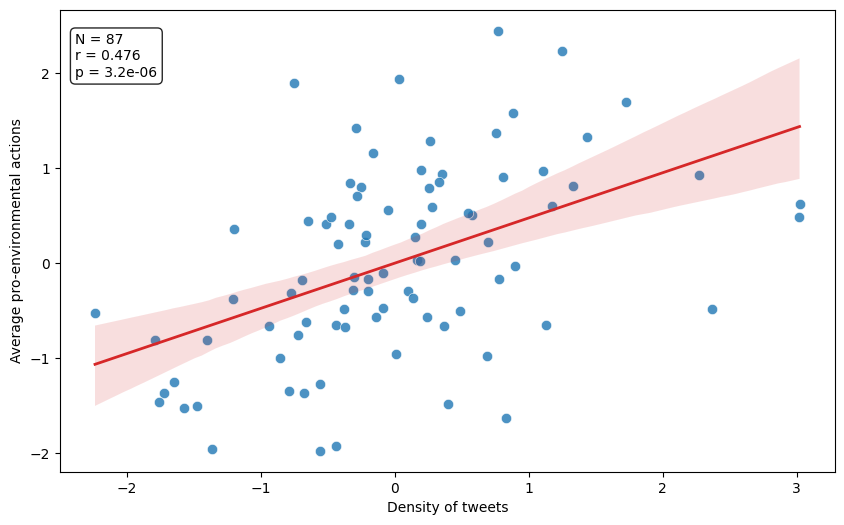}
        \caption{NUTS1}
        \label{fig:nuts1_scatter}
    \end{subfigure}
    \hfill
    \begin{subfigure}{0.49\linewidth}
        \centering
        \includegraphics[width=\linewidth]{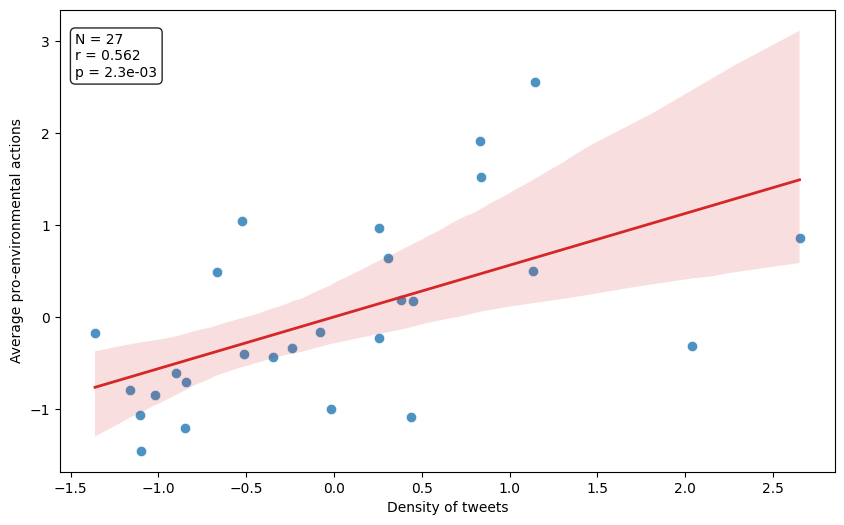}
        \caption{NUTS0}
        \label{fig:nuts0_scatter}
    \end{subfigure}
    \caption{Scatterplots of tweet density ($DenTweets$) and average pro-environmental actions ($ProEnv_{19}$) at different levels of territorial aggregation. Both variables are standardized (z-scores).}
\end{figure}

We further assess the robustness of the main findings by repeating the analysis at more aggregated territorial levels.
At the NUTS1 level, the positive association between tweet density and pro-environmental actions remains clear. As shown in Figure~\ref{fig:nuts1_scatter}, regions with higher levels of climate-related Twitter activity tend to report higher average levels of pro-environmental behavior, consistent with the baseline NUTS2 results.

A similar pattern emerges at the country level (NUTS0). Figure~\ref{fig:nuts0_scatter} shows a positive and statistically significant correlation between tweet density and pro-environmental actions, despite the smaller number of observations.
Across both aggregation levels, regression analyses in \STab{tab:regression_main_NUTS1} and \STab{tab:regression_main_NUTS0} controlling for education and social class confirm that the association between tweet density and offline pro-environmental behavior remains positive and robust, although confidence intervals widen at higher levels of aggregation due to reduced sample size.

Overall, the persistence of the association across spatial scales suggests that the link between online climate discourse and offline pro-environmental behavior is not driven by fine-grained regional variation alone, but reflects a broader and stable relationship.

Full regression results for these aggregation levels are provided in \STab{tab:regression_main_NUTS1} and \STab{tab:regression_main_NUTS0}.

\subsection{User analysis}

We additionally linked tweets to user identifiers, allowing us to compute the number of distinct users who posted at least one climate-related tweet in each region. We then repeated the same correlation and regression analyses performed for the density of tweets.
As shown in \SFig{fig:user_scatter}, and in \STab{tab:regression_user_main}, the density of users correlates positively and significantly with the average number of pro-environmental actions ($r = 0.499$, $p < 10^{-12}$), confirming the robustness of the association across different aggregation levels of the online activity signal, even when controlling for key socio-economic factors.

Overall, the results demonstrate that both the tweet-based and user-based measures lead to the same qualitative conclusions, confirming the stability of the relationship between online discourse and offline pro-environmental engagement.

\subsection{Believer tweets}

Since the Climate Change Twitter Dataset provides a stance classification (believer, neutral, or denier), we repeated the main analysis considering only tweets classified as “believer.” We computed the correlation (\SFig{fig:bel_scatter}) and performed the OLS regressions with Eurobarometer (\STab{tab:regression_main_bel}).

The results are consistent with those obtained using the total number of tweets, regardless of stance. The association between believer tweet density and pro-environmental behaviors remains positive and significant, but not stronger than that observed for total tweet density. 

Overall, these findings indicate that the predictive signal does not primarily depend on whether the content explicitly expresses belief in climate change. Rather, it is the total volume of climate-related discourse—reflecting the overall level of public attention and information exchange on the topic—that best predicts regional engagement in pro-environmental behaviors. This suggests that the informational intensity of the online debate, rather than its ideological orientation, plays a central role.

\subsection{Temporal sensitivity and outlier robustness}

We conducted additional checks to assess the stability of our results with respect to both the temporal window used to collect tweets and the inclusion of regions with limited data availability.

First, we tested the sensitivity of our findings to the choice of the starting date for tweet collection. For each starting point between 2015 and 2019, we re-estimated the OLS regression including tweet density and the main confounding factors. As shown in \SFig{fig:date_sens_EB}, the standardized coefficients remain stable in magnitude and statistical significance across different starting dates. In particular, tweet density consistently emerges as a positive and significant predictor of pro-environmental actions, while the coefficients of education and social class (or purchasing power) show only minor variation. The overall explanatory power of the model ($R^2$) also remains largely unchanged. These results indicate that our choice of January~1,~2017 as the baseline does not drive the observed association.

Second, we assessed robustness to outliers by progressively excluding regions with the lowest data availability and repeating the same regression analyses at each step. We separately removed regions with the fewest Eurobarometer respondents and those with the fewest geolocated tweets. As illustrated in \SFig{fig:EBremove1}, the estimated coefficients remain stable across most of the exclusion range, and only begin to fluctuate when a large number of regions is removed, reflecting reduced statistical power. The density of tweets consistently retains a positive and significant association with pro-environmental actions, confirming that the main results are not driven by a small subset of regions with extreme or sparse observations.

Taken together, these checks confirm that the relationship between online climate discourse and offline pro-environmental behavior is robust to both temporal aggregation choices and the presence of low-coverage regions.

% Replaced subsection as per instructions
\subsection{Robustness checks for NLP analyses}

We assess the robustness of the NLP-based results by examining how the Pearson correlation and the associated $p$-values vary across a wide range of quantile thresholds used to define each discourse dimension. All diagnostic figures referred to in this section are reported in the Appendix.

For \textit{activism}, the negative association with pro-environmental actions is stable across a broad range of high quantiles. Around the selected threshold (0.95), both the correlation and its statistical significance vary smoothly, indicating that the result is not driven by a knife-edge cutoff. In addition, a secondary local minimum is observed around the 0.80 quantile, where the association remains negative and statistically significant (see \SFig{fig:corr_plot_act}).

For \textit{social support}, the correlation profile displays a clear and well-defined minimum around the selected threshold (0.97). Both the Pearson correlation and the corresponding $p$-values are stable in the neighborhood of this quantile, and no competing local extrema emerge at lower thresholds (see \SFig{fig:corr_plot_supp}).

In contrast, for \textit{knowledge exchange} no quantile yields a stable or meaningful association with pro-environmental actions. Across the entire range of tested thresholds, correlations remain close to zero and statistically insignificant, supporting the decision not to construct a regional predictor for this dimension (see \SFig{fig:corr_plot_know}).

\section{Discussion}

We find a clear positive association between the density of climate-related tweets and the average adoption of pro-environmental behaviors across European regions. This association remains significant when controlling for multiple sets of confounding factors and across robustness checks. This provides a rare cross-regional link between large-scale online climate discourse and harmonized survey-based measures of offline pro-environmental behavior. The NLP extension further shows that different forms of online engagement are not equivalent: activism- and support-oriented discourse are negatively associated with offline actions, while knowledge exchange shows no clear relationship.

A natural interpretation of the positive association between tweet density and offline actions is that it reflects differences in public attention across regions. In this view, denser climate discourse indicates that climate change is more salient in everyday communication, which can support coordination by making pro-environmental actions more cognitively available and socially discussable.
In this sense, online attention can be interpreted as a proxy for a communication environment in which coordination around pro-environmental practices is more likely to emerge, without implying that attention itself causes adoption.
At the same time, several alternative mechanisms are consistent with the same empirical pattern. 
Tweet density may also capture broader regional characteristics that jointly shape both online communication and offline behavior, including education and socio-economic conditions, as well as other unobserved factors.
For this reason, we interpret tweet density as a high-frequency proxy of regional attention to climate issues rather than as a causal driver of pro-environmental action.

In addition, NLP results indicate that volume alone can mask differences in the social intent of climate talk. We therefore treat tweet density as an attention-related signal, and use composition-based NLP indicators to interpret how different communicative functions relate to offline action.
The negative associations observed for activism- and social support–related discourse may appear counterintuitive, but they are consistent with the idea that mobilization-oriented and supportive communication can play different roles across contexts.
In our NLP measures, \textit{activism} captures mobilization-oriented expressions (e.g., calls to action or participation intent), whereas \textit{social support} captures language offering encouragement, empathy, or practical help.
A first interpretation is compensatory. Where pro-environmental actions are already widespread and socially normalized, there may be less need for calls to action or public encouragement. Conversely, higher prevalence of mobilization- or support-oriented language may emerge where offline adoption is lower, either because barriers are higher or because practices are less consolidated.
A second interpretation is that activism and support language can reflect low-cost or primarily expressive engagement that does not necessarily translate into measurable offline action. In this view, people may express support and share calls to action online, but whether this leads to offline behavior still depends on practical constraints and opportunities. This reading is consistent with review evidence emphasizing that links between social media engagement and behavioral change are mixed and context-dependent \citep{gaupp2024climate}.
These interpretations are plausible but necessarily speculative given the observational design, and we cannot distinguish between them here. Nonetheless, the negative associations highlight that mobilization- and support-oriented discourse should not be treated as direct proxies for behavioral adoption. While activism captures mobilization intent and calls to action, support reflects encouragement and coping-oriented exchanges, and both may be most visible where attention is high but adoption remains difficult.
Finally, we find no clear association for knowledge exchange. This does not imply that information is irrelevant, but rather that, at this level of aggregation, knowledge-oriented content does not help explain cross-regional differences in reported actions, either because it is widespread across regions or because it is only loosely connected to the specific actions captured in the survey.

These findings have two main implications. First, combining digital traces with harmonized survey data helps characterize cross-regional differences in pro-environmental action in a way that is difficult to achieve with either source alone. In particular, tweet density can be read as an attention-related signal that complements standard socio-economic indicators.
Second, the NLP results caution against treating online engagement as a single construct. Regions with similar levels of climate attention can differ in the social intent of their climate conversations, and these differences relate to offline action in different directions.
From a practical perspective, the results suggest that spikes in activism- or support-oriented discourse should not automatically be read as evidence of higher behavioral adoption. Instead, they may flag contexts where climate issues are salient but adoption is more difficult, and where interventions that reduce barriers or translate attention into concrete opportunities could be most valuable.

Some limitations should be kept in mind when interpreting these results. First, the analysis is observational and cross-sectional at the regional level. Even with extensive controls, unobserved factors may jointly shape both online discourse and offline behavior, and the association may reflect bidirectional relationships, common underlying drivers, or both. For this reason, the results should not be interpreted causally.

Second, our outcomes and predictors are aggregated at the regional scale. This raises standard concerns about ecological inference: associations observed across regions do not necessarily describe individual-level relationships, and regional averages can mask within-region heterogeneity.

Third, both data sources have measurement constraints. Pro-environmental actions are self-reported, which can introduce social desirability and recall bias. Twitter coverage is uneven across regions and populations, and Twitter users are not representative of the general public. At the same time, some Twitter-based indicators, especially those capturing attention, can align with survey-based measures of public concern and can therefore be used as reasonably informative proxies of attention at aggregate levels \citep{crispino2024people}. Geolocation relies on metadata and user-declared profile locations for a large share of tweets, which introduces uncertainty in mapping tweets to regions, despite validation efforts \citep{effrosynidis2022climate}.

Fourth, our analysis focuses on Twitter and on European regions only. Patterns may differ on other platforms with different user bases and affordances, and in non-European institutional and cultural contexts. Moreover, our analysis is anchored to a single survey snapshot from 2019 and our Twitter signal is measured over 2017--2019, so we cannot assess whether the same associations hold in earlier or later phases of the online climate debate. While robustness checks indicate that the tweet-density association is stable when the tweet collection window is moved closer to the survey period, future work should assess whether similar online signals replicate across different survey waves and years.

Finally, the NLP-based dimensions depend on classifier performance and on the thresholding choices used to construct regional shares, even though we performed robustness checks across alternative thresholds. Misclassification can propagate into regional aggregates, and our indicators cover only a limited subset of the full complexity of climate communication. For activism in particular, the classifier is topic-agnostic and detects generic mobilization language, which may not always correspond to climate-specific collective action.

\section{Conclusion}

This paper provides a cross-regional analysis linking online climate discourse on Twitter (2017--2019) to harmonized survey-based measures of offline pro-environmental actions across European regions, using the 2019 Special Eurobarometer. We find a robust positive association between tweet density and the average number of reported actions, which remains stable across alternative specifications and robustness checks, including models with Eurostat-based controls, consistent with the view that online attention contains informative signals about regional behavioral patterns.

At the same time, the NLP extension highlights heterogeneity in online engagement. Activism- and support-oriented language is negatively associated with reported actions, while knowledge exchange shows no clear relationship. Together, these findings reinforce a cautious interpretation of social media indicators: online discourse can be informative for monitoring and comparative analysis, but its meaning depends on the social intent of communication and should not be read causally.

Future work should test the stability of these patterns across time and settings, for example using multiple survey waves and longer observation windows, and assess whether similar signals replicate on other platforms (e.g. Reddit, Bluesky). It would also be valuable to apply the same framework with alternative survey sources that cover other parts of the world (for example Gallup-based measures in the United States or global public opinion series), to assess generalizability beyond Europe. Finally, future research could validate the findings against alternative offline outcomes where available (for example energy use, mobility, or purchasing proxies) and expand the discourse decomposition with additional dimensions and domain-specific validation of NLP classifiers.

More broadly, our results support the use of social media signals as complementary, attention-based indicators for monitoring where pro-environmental adoption may be harder, helping target and evaluate interventions while remaining cautious about causal interpretation.

\clearpage

% appendix

\appendix
\renewcommand{\thesection}{A}
\renewcommand{\thefigure}{A.\arabic{figure}}
\renewcommand{\thetable}{A.\arabic{table}}
\setcounter{figure}{0}
\setcounter{table}{0}
\section*{Appendix}
\addcontentsline{toc}{section}{Appendix}

\subsection*{List of possible pro-environmental actions}
The Eurobarometer survey includes a list of possible pro-environmental actions that participants can choose from to answer to the question "Have you done any of the following in the past six months?". The list includes the following actions:
\begin{enumerate}[noitemsep]
    \item Chosen a more environmentally-friendly way of travelling (walk, bicycle,
    public transport, electric car)
    \item Avoided buying over-packaged products
    \item Avoided single-use plastic goods other than plastic bags (e.g. plastic
    cutlery, cups, plates, etc.) or bought reusable plastic products
    \item Separated most of your waste for recycling 
    \item Cut down your water consumption
    \item Cut down your energy consumption (e.g. by turning down air conditioning
    or heating, not leaving appliances on stand-by, buying energy-efficient
    appliances)
    \item Bought products marked with an environmental label 
    \item Bought local products 
    \item Used your car less by avoiding unnecessary trips, working from home
    (teleworking), etc.
    \item Joined a demonstration, attended a workshop, taken part in an activity
    (e.g. a collective beach or park cleanup) (N)
    \item Changed your diet to more sustainable food (N) 
    \item Spoken to others about environmental issues (N) 
    \item Bought second-hand products (e.g. clothes or electronics) instead of new
    ones (N)
    \item Repaired a product instead of replacing it (N) 
    \item None (SPONTANEOUS) 
    \item DK
\end{enumerate}
The actions marked with (N) are considered "new" actions, included in the survey for the first time in 2019. DK means "Don't know".

\subsection*{Additional figures}
\iftrue
\begin{figure}[h]
    \centering
    \includegraphics[width=0.53\textwidth]{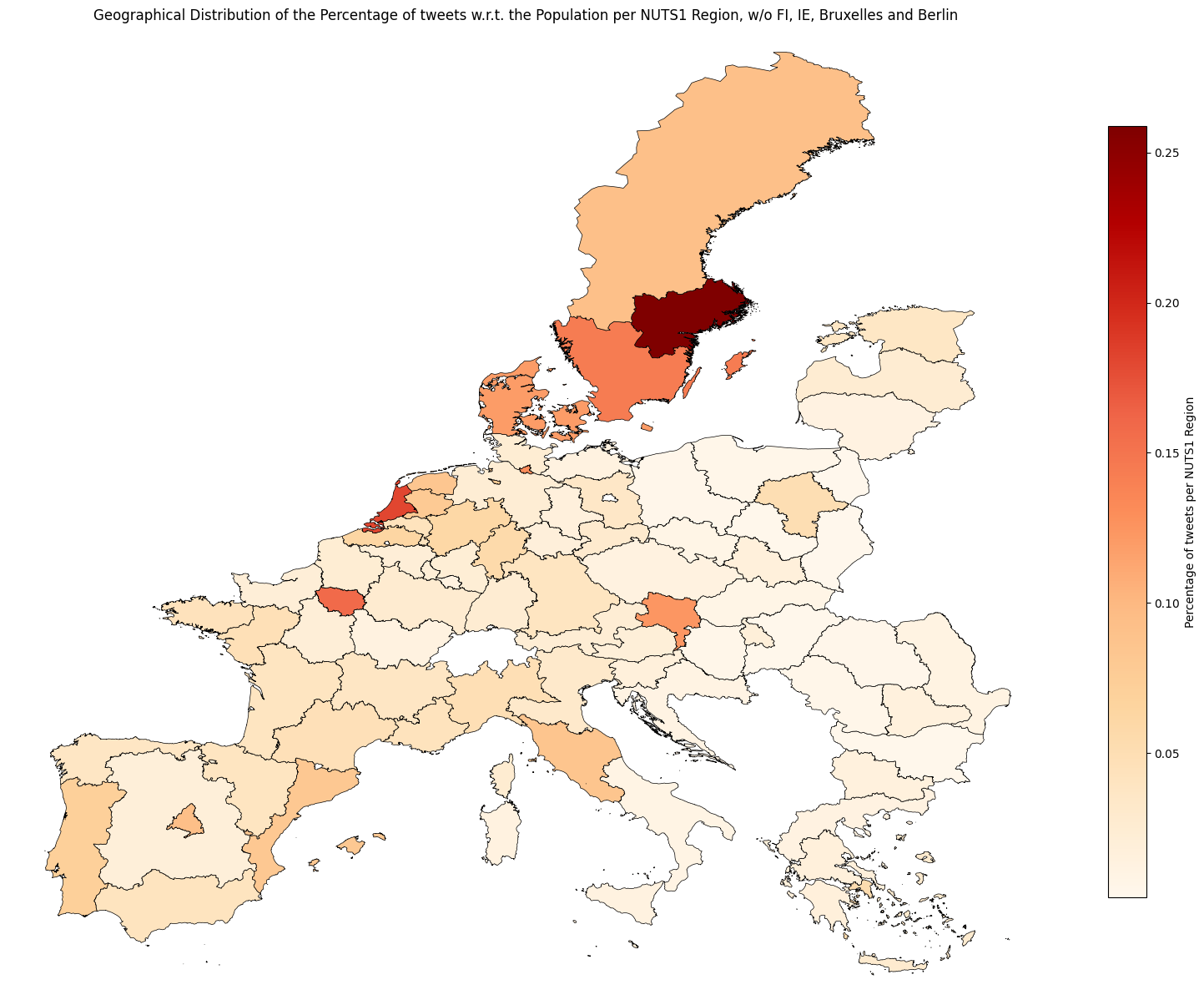}
    \caption{Fraction of tweets w.r.t. the population in the NUTS1 regions in the EU. Finland (1.25\%) and Ireland (0.5\%), Brussels (1.26\%), and Berlin (0.50\%) are not shown as out of scale. }
    \label{fig:map_perc_nuts1_X2}
\end{figure}

\begin{figure}[h]
    \centering
    \begin{subfigure}{0.494\textwidth}
        \centering
        \includegraphics[width=\textwidth]{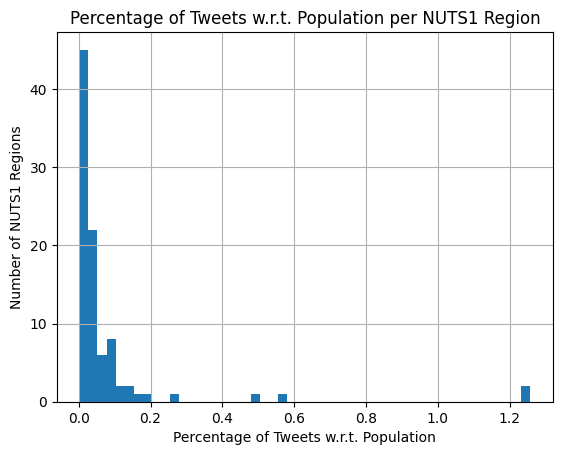}
        \caption{Linear-linear scale}
        \label{fig:dist_perc_nuts1_lin}
    \end{subfigure}
    \begin{subfigure}{0.48\textwidth}
        \centering
        \includegraphics[width=\textwidth]{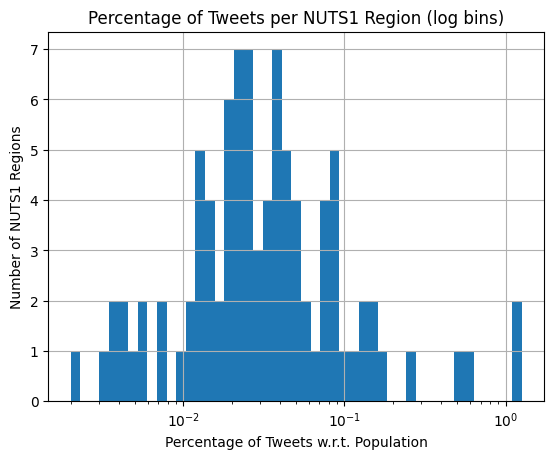}
        \caption{Log-linear scale}
        \label{fig:dist_perc_nuts1_log}
    \end{subfigure}
    \caption{Distribution of the percentage of tweets w.r.t. the population in each NUTS1 region. On the left, in a linear scale. On the right, in a logarithmic scale with logarithmic bins. Mean = 0.08\% and Standard Deviation = 0.195\% considering all the regions.}
    \label{fig:dist_perc_nuts1}
\end{figure}
\fi

\begin{figure}
    \centering
    \includegraphics[width=0.9\linewidth]{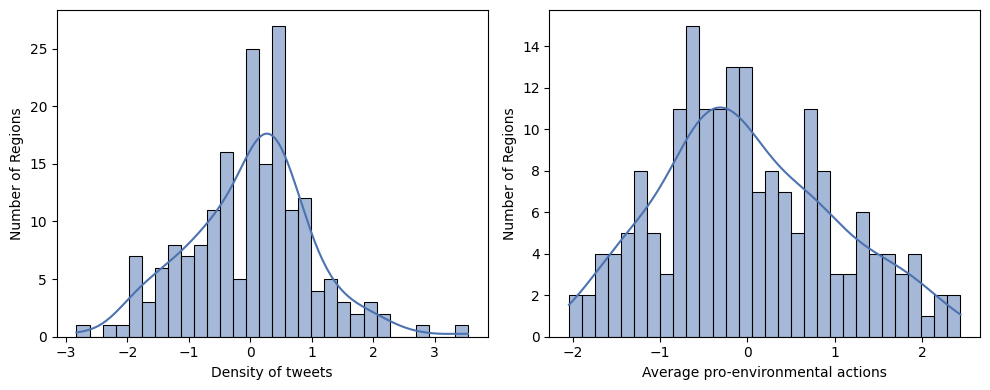}
    \caption{Distributions of standardized variables. Left: Density of tweets per capita (log). Right: Average pro-environmental actions per participant. All variables are z-scored across regions.}
    \label{fig:dist_tweet_proenv}
\end{figure}

\begin{figure}
    \centering
    \includegraphics[width=0.9\linewidth]{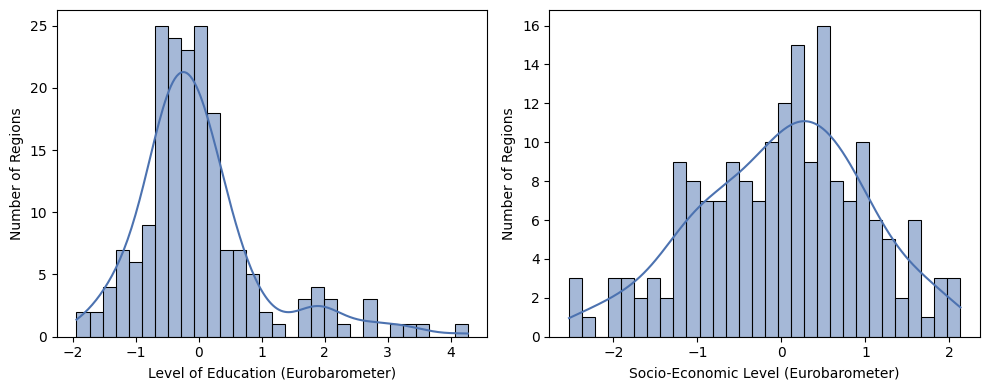}
    \caption{Distributions of standardized Eurobarometer covariates. Left: Level of Education (log years of full-time education). Right: Socio-Economic Level (self-reported social class). All variables are z-scored across regions.}
    \label{fig:dist_EBmain}
\end{figure}
\begin{figure}
    \centering
    \includegraphics[width=0.9\linewidth]{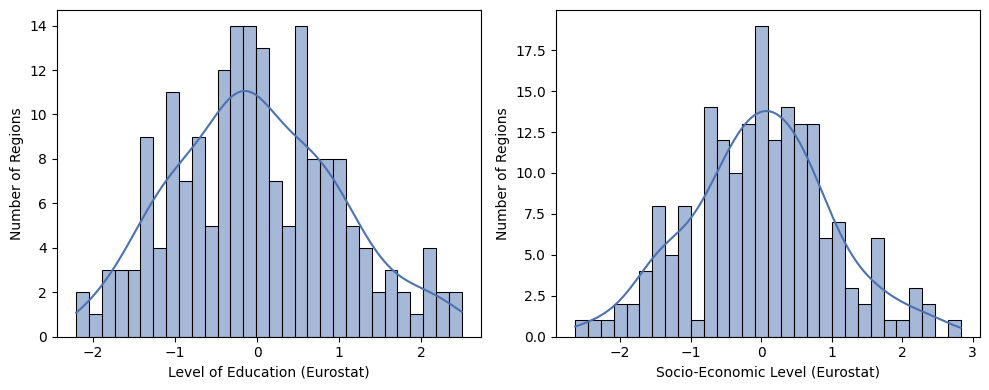}
    \caption{Distributions of standardized Eurostat covariates. Left: Level of Education (fraction with tertiary education). Right: Socio-Economic Level (PPS per inhabitant, log). All variables are z-scored across regions.}
    \label{fig:dist_ESmain}
\end{figure}

\begin{figure}[h]
    \centering
    \begin{subfigure}[b]{0.48\linewidth}
    \includegraphics[width=\linewidth]{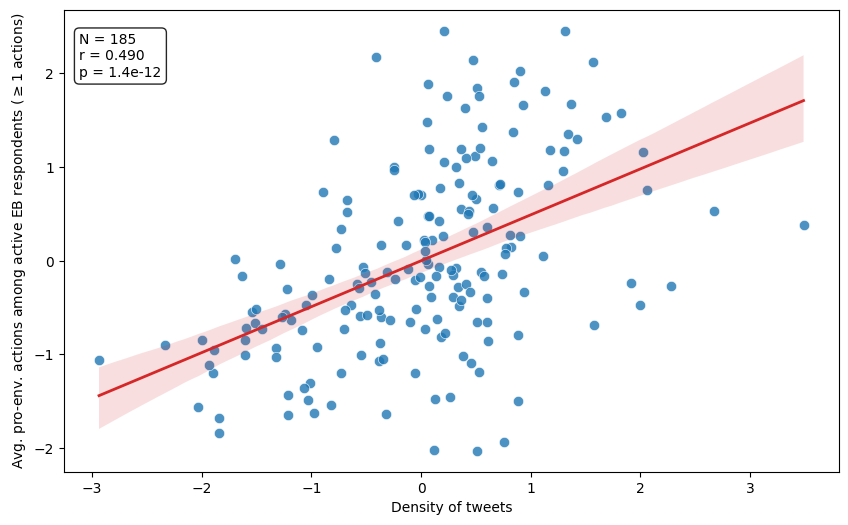}
    \caption{Intensive: average number of pro-environmental actions among active respondents ($\geq$1 action).}
    \label{fig:int_ge1}
    \end{subfigure}
    \hfill
    \begin{subfigure}[b]{0.48\linewidth}
    \includegraphics[width=\linewidth]{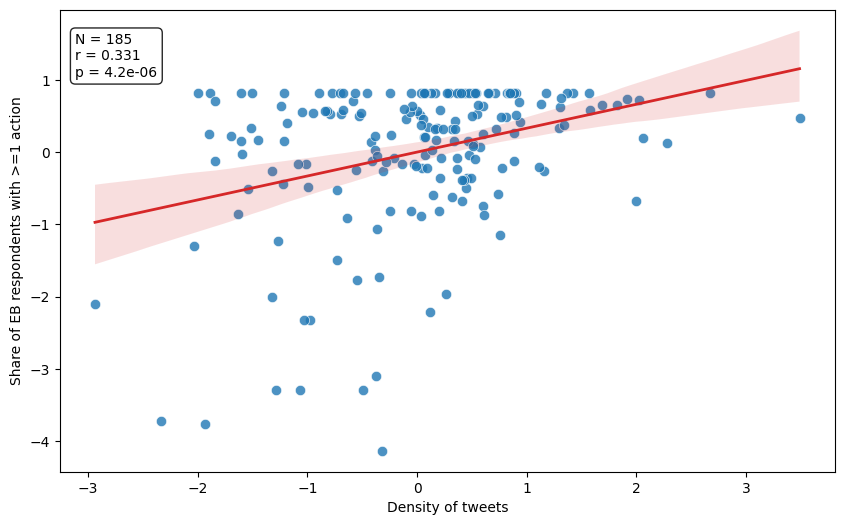}
    \caption{Extensive: share of Eurobarometer respondents with at least one action.}
    \label{fig:ext_ge1}
    \end{subfigure}
    \caption{Correlation between tweet density and pro-environmental behavior for active respondents (threshold $\geq$1). Both the intensive (a) and extensive (b) components are positively and significantly correlated with tweet density ($r=0.49$ and $r=0.33$, respectively; $p<0.001$).}
    \label{fig:int_ext_ge1}
\end{figure}

\begin{figure}[h]
    \centering
    \begin{subfigure}[b]{0.32\linewidth}
        \centering
        \includegraphics[width=\linewidth]{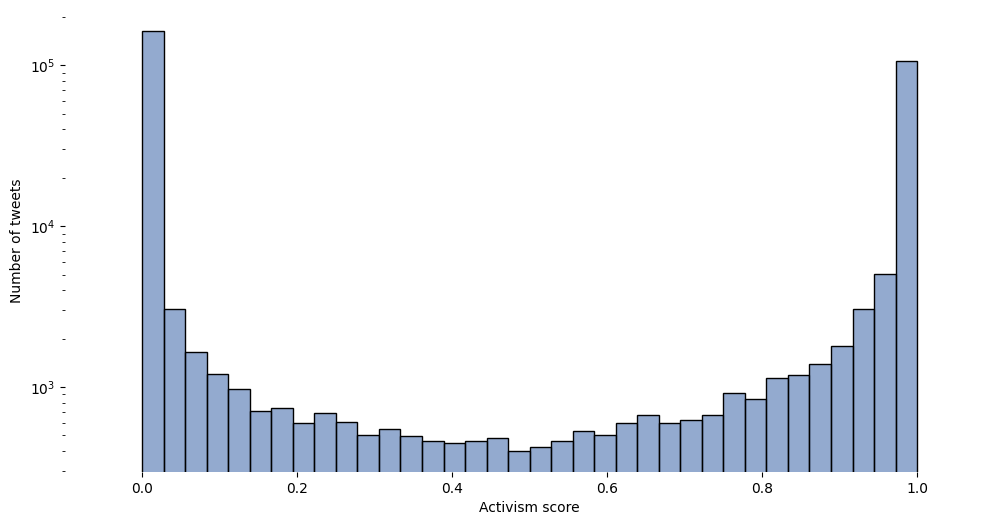}
        \caption{\centering Distribution of the activism score across all regions.}
        \label{fig:act_dist}
    \end{subfigure}
    \hfill
    \begin{subfigure}[b]{0.32\linewidth}
        \centering
        \includegraphics[width=\linewidth]{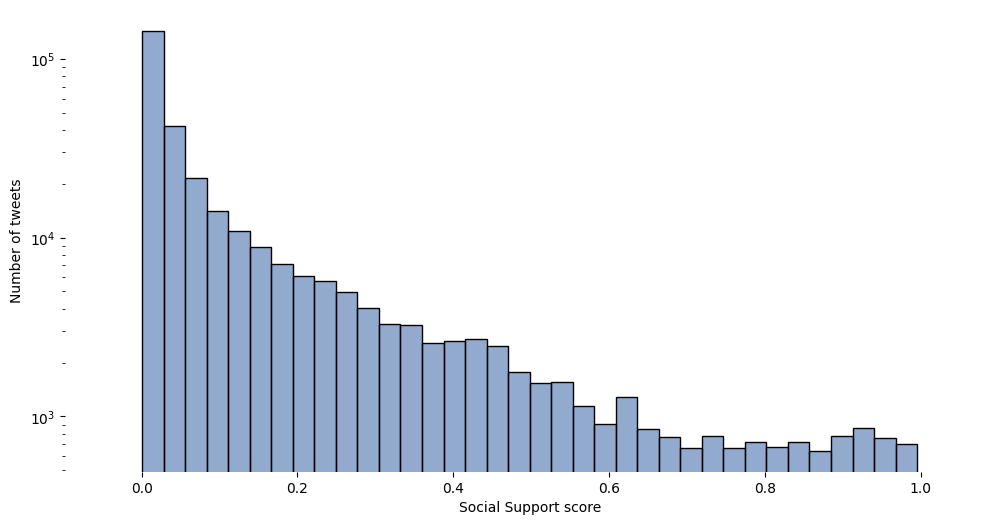}
        \caption{\centering Distribution of the social support score across all regions.}
        \label{fig:supp_dist}
    \end{subfigure}
    \hfill
    \begin{subfigure}[b]{0.32\linewidth}
        \centering
        \includegraphics[width=\linewidth]{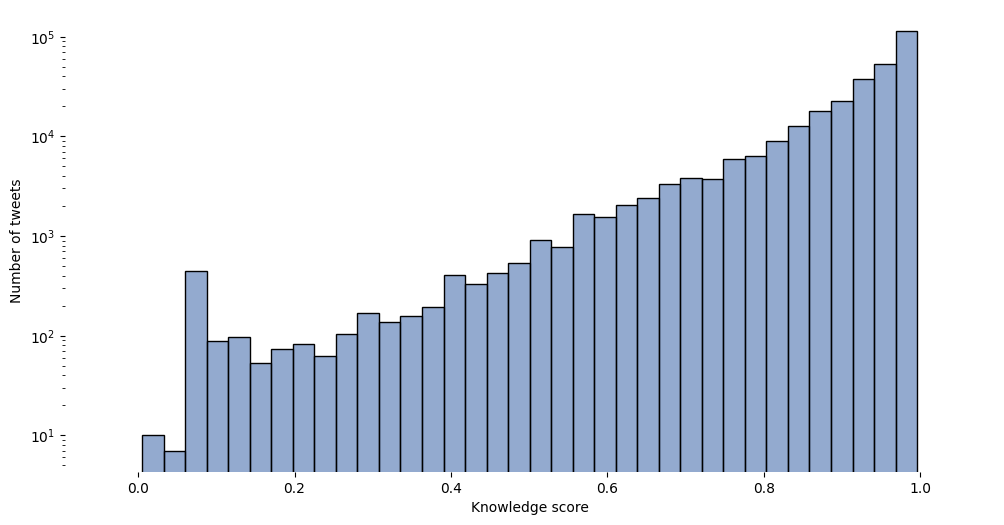}
        \caption{\centering Distribution of the knowledge score across all regions.}
        \label{fig:know_dist}
    \end{subfigure}
    \caption{Distributions of the activism, social support, and knowledge scores across all regions.}
    \label{fig:dim_dist}
\end{figure}

\begin{figure}[h]
    \centering
    \includegraphics[width=0.75\linewidth]{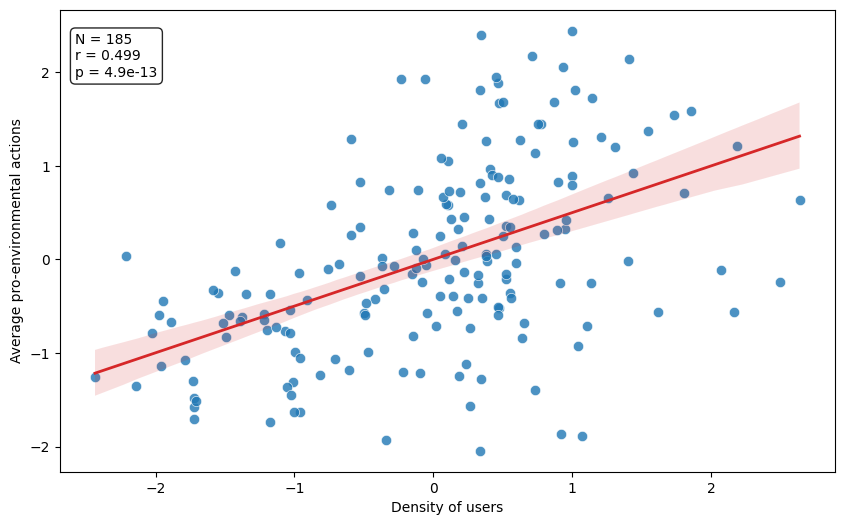}
    \caption{Scatterplot of $DenUsers$, logarithm of the density of users per region, normalized by the regional population, and $ProEnv_{19}$, the average number of self-reported pro-environmental actions (from Eurobarometer 2019) per capita. Z-scores computed for both variables. Pearson’s correlation coefficient $r=0.499$, $p=4.9\times10^{-13}$, with 185 regions considered.}
    \label{fig:user_scatter}
\end{figure}

\begin{figure}[h]
    \centering
    \includegraphics[width=0.85\linewidth]{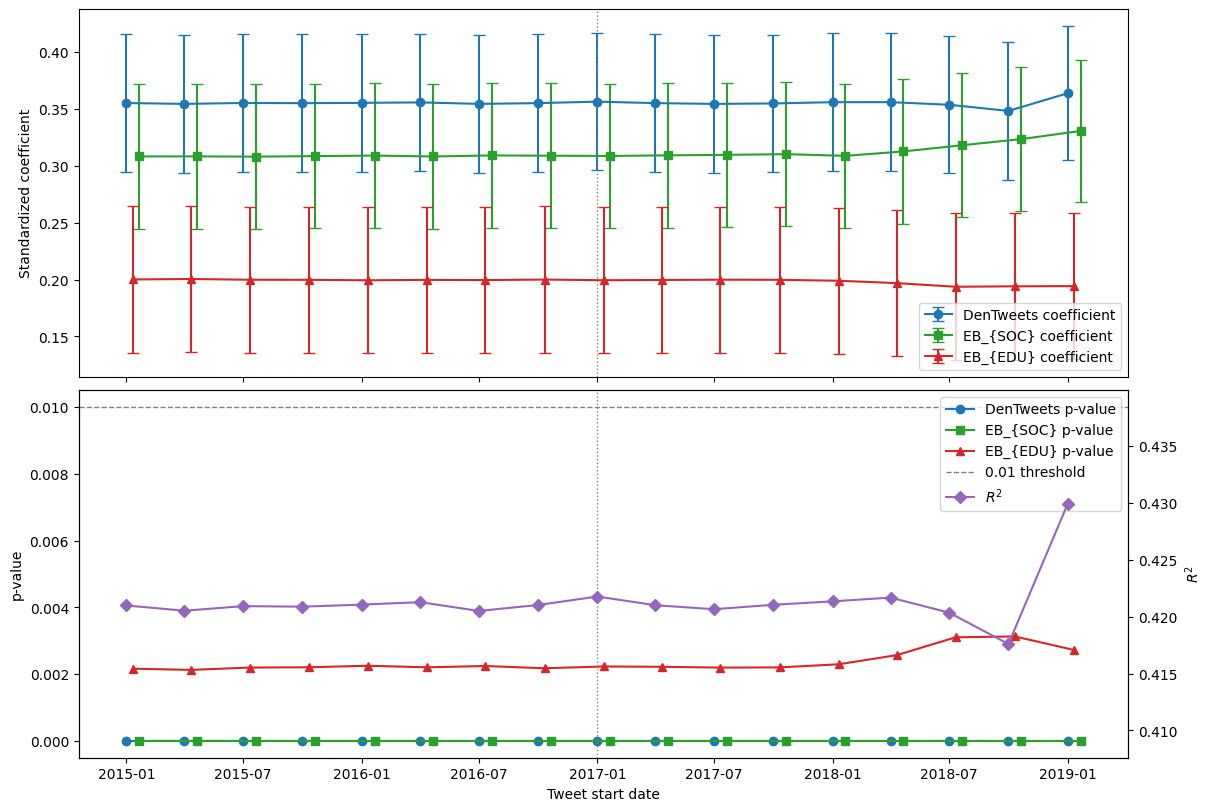}
    \caption{Sensitivity analysis of the OLS regression coefficients with Eurobarometer controls across different tweet collection start dates. The upper panel shows standardized coefficients with their standard errors, and the lower panel shows the corresponding $p$-values and model $R^2$. The vertical dashed line marks the chosen start date of 01/01/2017. The stability of coefficients, standard errors, and significance levels confirms the robustness of the results to temporal window variations.}
    \label{fig:date_sens_EB}
\end{figure}

\begin{figure}[h]
    \centering
    \includegraphics[width=0.85\linewidth]{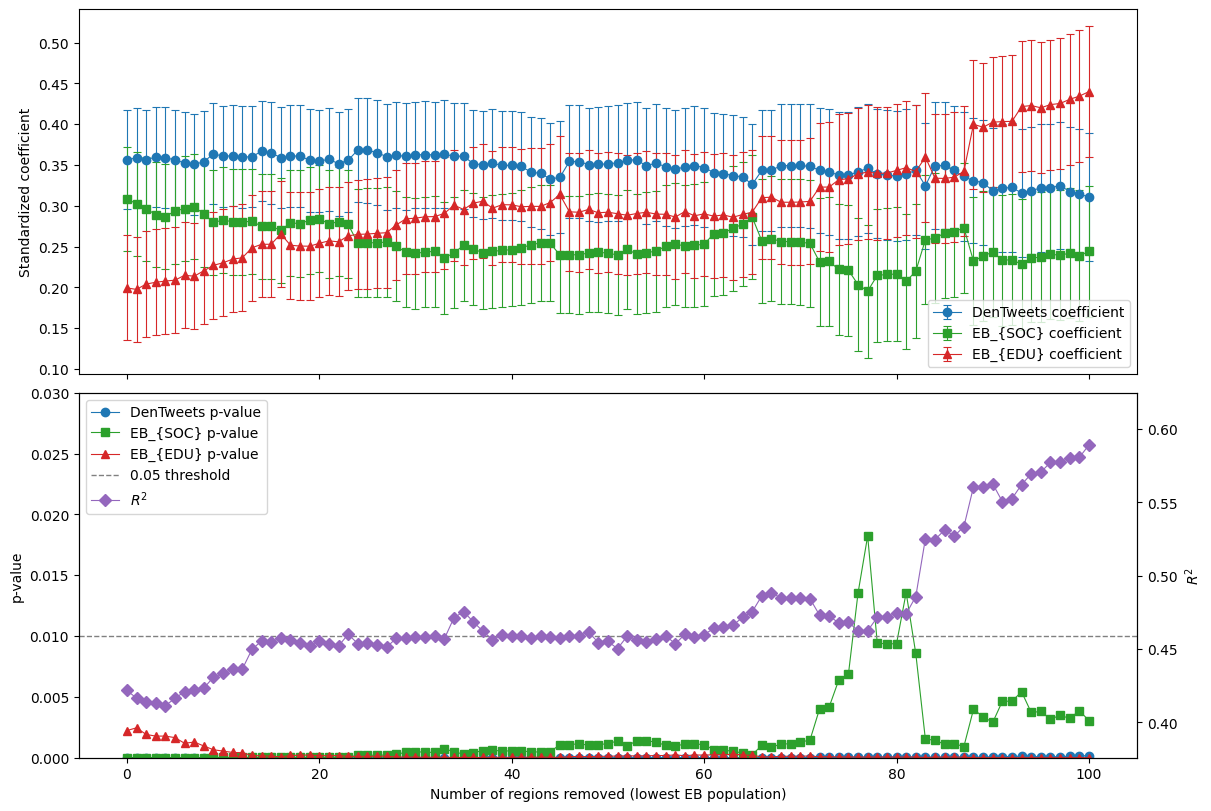}
    \caption{Sensitivity analysis of the OLS regression with Eurobarometer controls when progressively removing regions with the fewest Eurobarometer respondents. The upper panel shows standardized coefficients with standard errors, and the lower panel shows the corresponding $p$-values and model $R^2$. The results remain stable up to the removal of roughly 70 regions, confirming the robustness of the association between tweet density and pro-environmental behaviors.}
    \label{fig:EBremove1}
\end{figure}

\begin{figure}[h]
    \centering
    \includegraphics[width=0.85\linewidth]{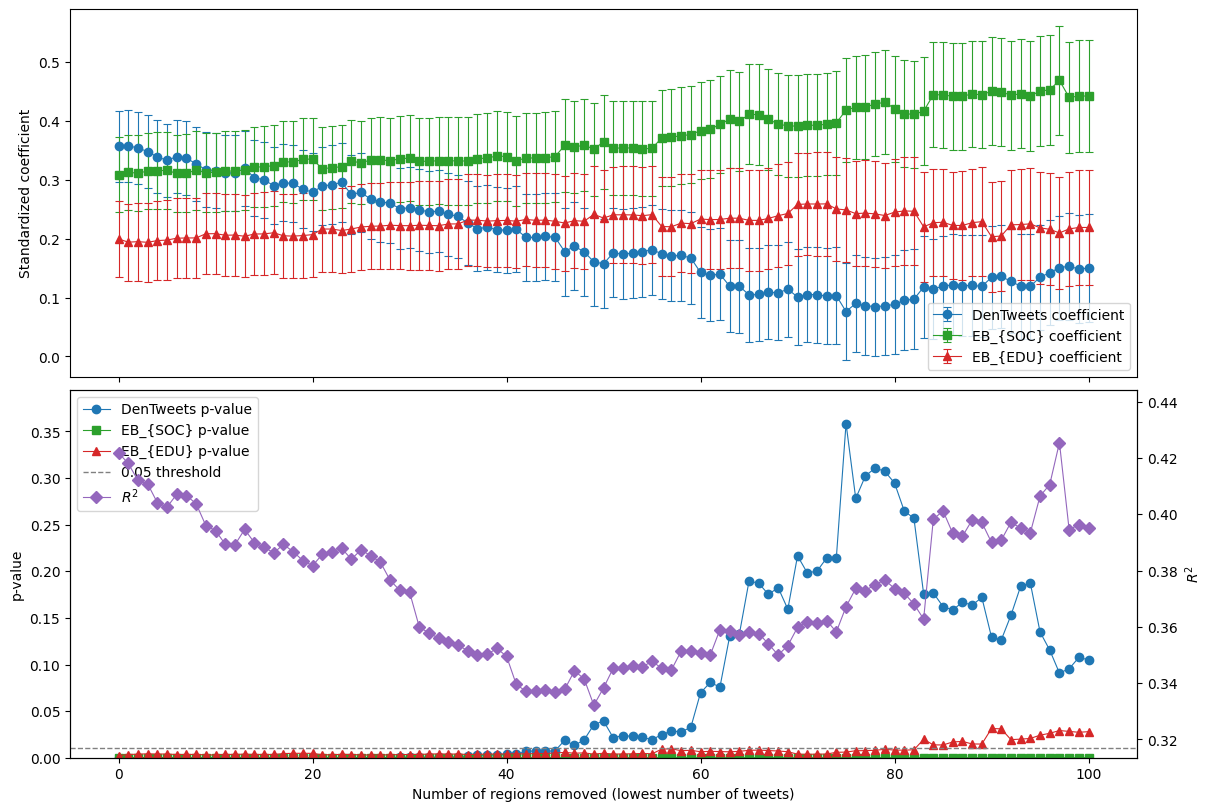}
    \caption{Sensitivity analysis of the OLS regression with Eurobarometer controls when progressively removing regions with the fewest geolocated tweets. The upper panel shows standardized coefficients with standard errors, and the lower panel the corresponding $p$-values and model $R^2$. The coefficients remain stable until approximately 60 regions are removed, after which variance increases due to smaller sample size.}
    \label{fig:EBremove2}
\end{figure}

\begin{figure}[h]
    \centering
    \includegraphics[width=0.75\linewidth]{figures/bel_scatter.png}
    \caption{Scatterplot of $BelTweets$, logarithm of the density of believer tweets per region normalized by population, and $ProEnv_{19}$, average number of self-reported pro-environmental actions (from Eurobarometer 2019) per capita. Z-scores computed for both variables. Pearson’s correlation coefficient $r = 0.501$, $p = 3.8 \times 10^{-13}$, 185 regions considered.}
    \label{fig:bel_scatter}
\end{figure}

\begin{figure}[h]
    \centering
    \includegraphics[width=0.75\linewidth]{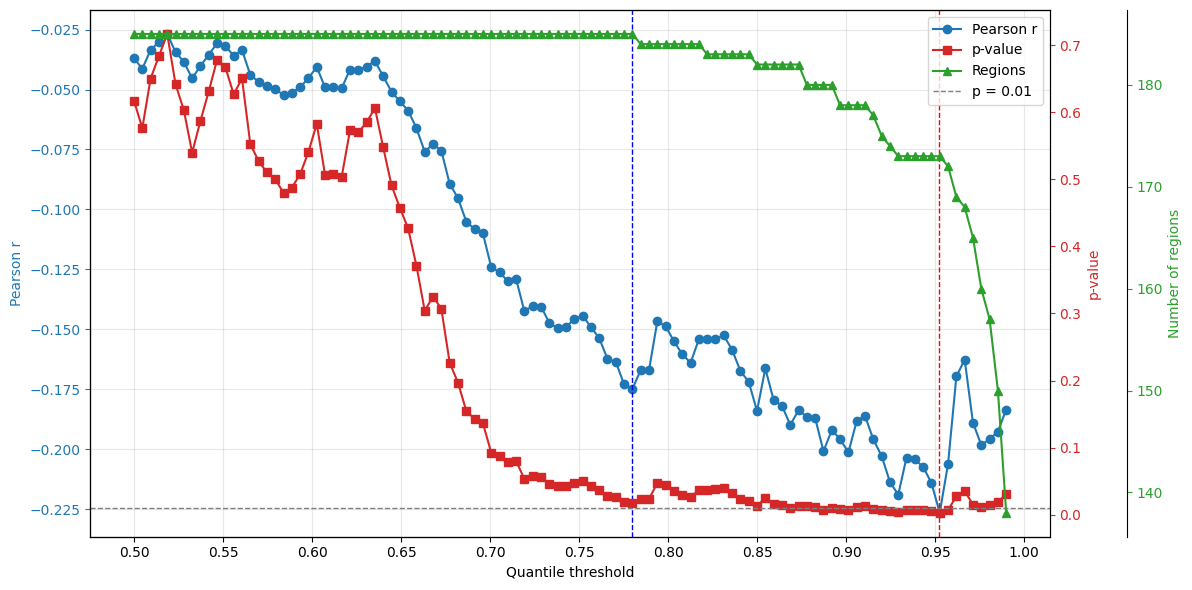}
    \caption{Sensitivity of the Pearson correlation between activism-related tweets and pro-environmental actions to the quantile threshold used to define activist tweets. The plot shows the correlation, corresponding $p$-values, and the number of regions with nonzero values as functions of the threshold.}
    \label{fig:corr_plot_act}
\end{figure}

\begin{figure}[h]
    \centering
    \includegraphics[width=0.75\linewidth]{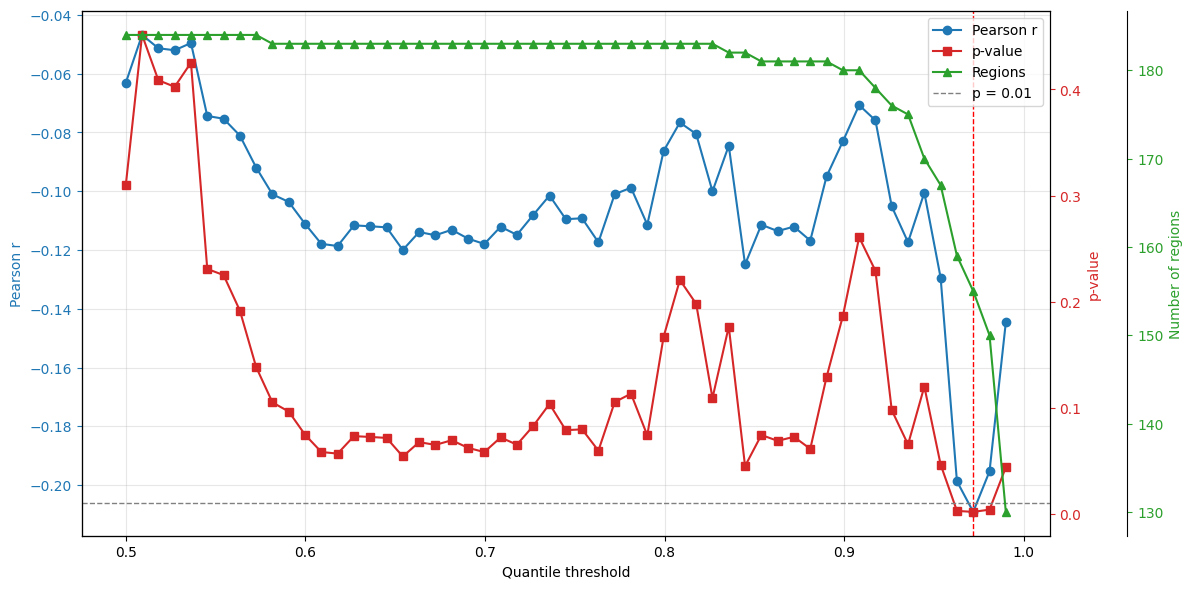}
    \caption{Sensitivity of the Pearson correlation between social support-related tweets and pro-environmental actions to the quantile threshold used to define support tweets. The plot shows the correlation, corresponding $p$-values, and the number of regions with nonzero values as functions of the threshold.}
    \label{fig:corr_plot_supp}
\end{figure}

\begin{figure}[h]
    \centering
    \includegraphics[width=0.75\linewidth]{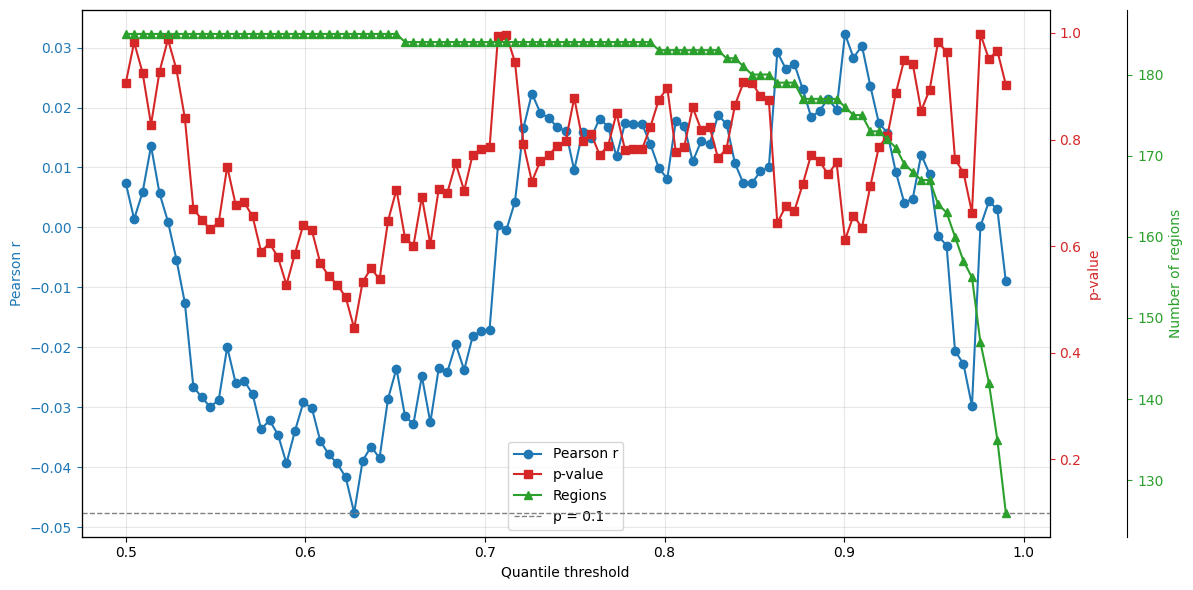}
    \caption{Sensitivity of the Pearson correlation between knowledge-related tweets and pro-environmental actions to the quantile threshold used to define knowledge tweets. No threshold yields a stable or significant association.}
    \label{fig:corr_plot_know}
\end{figure}

\newpage
\newpage
\cleardoublepage
\subsection*{Additional tables}

\begin{table}[h]
    \begin{tabular}{lccccc}
        \toprule
        \textbf{$ProEnv_{19}$} & (1) & (2) & (3) & (4) & (5) \\
        \midrule
        \textbf{$DenTweets$} &  &  &  &  & $0.318^{***}$ \\
         &  &  &  &  & (0.064) \\

        \textbf{$EB_{EDU}$} &  &  &  &  & $0.160^{**}$ \\
         &  &  &  &  & (0.064) \\

        \textbf{$EB_{SOC}$} &  &  &  &  & $0.179^{**}$ \\
         &  &  &  &  & (0.072) \\

        \textbf{$EB_{ECON}$} & $0.433^{***}$ &  &  &  & $0.142^{**}$ \\
         & (0.067) &  &  &  & (0.063) \\

        \textbf{$EB_{CITY}$} &  & $0.087$ &  &  & $-0.128^{**}$ \\
         &  & (0.074) &  &  & (0.059) \\

        \textbf{$EB_{INT}$} &  &  & $0.544^{***}$ &  & $0.167^{**}$ \\
         &  &  & (0.062) &  & (0.076) \\

        \textbf{$EB_{POL}$} &  &  &  & $-0.318^{***}$ & $-0.081$ \\
         &  &  &  & (0.070) & (0.059) \\

        \midrule
        \# regions & 185 & 185 & 185 & 185 & 185 \\
        $R^2$ & 0.187 & 0.008 & 0.296 & 0.101 & 0.488 \\
        \bottomrule
    \end{tabular}

    \caption{Regression results for the average number of pro-environmental actions ($ProEnv_{19}$). The main predictor is $DenTweets_{19}$, while confounding factors are $EB_{EDU}$, $EB_{SOC}$, $EB_{ECON}$, $EB_{CITY}$, $EB_{INT}$, and $EB_{POL}$, which are the average years of full-time education, social class of Eurobarometer respondents, fraction of respondents reporting difficulties in paying bills, type of area of residence, frequency of internet use, and political orientation respectively, all computed from the Eurobarometer. For all variables, we computed the z-score. Reported coefficients are shown with standard errors in parentheses. Regions considered at NUTS2 level. * p $<$ 0.1, ** p $<$ 0.05, *** p $<$ 0.01. }
    \label{tab:regression_addEB}
\end{table}

\begin{table}[h]
    \begin{tabular}{lccccccc}
        \toprule
        \textbf{$ProEnv_{19}$} & (1) & (2) & (3) & (4) & (5) & (6) & (7)\\
        \midrule
        \textbf{$DenTweets$} & $0.476^{***}$ &  &  & $0.367^{***}$ & $0.394^{***}$ &  & $0.337^{***}$\\
         & (0.095) &  &  & (0.107) & (0.098) &  & (0.106)\\

        \textbf{$EB_{EDU}$} &  & $0.402^{***}$ &  & $0.225^{**}$ &  & $0.291^{***}$ & $0.150$\\
         &  & (0.099) &  & (0.107) &  & (0.108) & (0.112)\\

        \textbf{$EB_{SOC}$} &  &  & $0.380^{***}$ &  & $0.251^{**}$ & $0.251^{**}$ & $0.204^{*}$\\
         &  &  & (0.100) &  & (0.098) & (0.108) & (0.104)\\

        \midrule
        Number of regions & 87 & 87 & 87 & 87 & 87 & 87 & 87\\
        $R^2$ & 0.226 & 0.162 & 0.144 & 0.265 & 0.283 & 0.213 & 0.298\\
        \bottomrule
    \end{tabular}

    \caption{Regression results for the average number of pro-environmental actions ($ProEnv_{19}$). The main predictor is the number of tweets normalized by the regional population ($DenTweets$), while confounding factors include the average number of years of full-time education ($EB_{EDU}$) and the average social class ($EB_{SOC}$), both computed from the Eurobarometer. For all variables, the z-score is considered. Reported coefficients are shown with standard errors in parentheses. Regions considered at NUTS1 level. * p $<$ 0.1, ** p $<$ 0.05, *** p $<$ 0.01. }
    \label{tab:regression_main_NUTS1}
\end{table}

\begin{table}[h]
    \begin{tabular}{lccccccc}
        \toprule
        \textbf{$ProEnv_{19}$} & (1) & (2) & (3) & (4) & (5) & (6) & (7)\\
        \midrule
        \textbf{$DenTweets$} & $0.562^{***}$ &  &  & $0.312^{**}$ & $0.417^{***}$ &  & $0.295^{**}$\\
         & (0.165) &  &  & (0.146) & (0.135) &  & (0.130)\\

        \textbf{$EB_{EDU}$} &  & $0.711^{***}$ &  & $0.575^{***}$ &  & $0.505^{***}$ & $0.385^{**}$\\
         &  & (0.141) &  & (0.146) &  & (0.148) & (0.147)\\

        \textbf{$EB_{SOC}$} &  &  & $0.657^{***}$ &  & $0.547^{***}$ & $0.392^{**}$ & $0.377^{**}$\\
         &  &  & (0.151) &  & (0.135) & (0.148) & (0.137)\\

        \midrule
        Number of regions & 27 & 27 & 27 & 27 & 27 & 27 & 27\\
        $R^2$ & 0.316 & 0.505 & 0.432 & 0.584 & 0.594 & 0.617 & 0.687\\
        \bottomrule
    \end{tabular}

    \caption{Regression results for the average number of pro-environmental actions ($ProEnv_{19}$). The main predictor is the number of tweets normalized by the regional population ($DenTweets$), while confounding factors include the average number of years of full-time education ($EB_{EDU}$) and the average social class ($EB_{SOC}$), both computed from the Eurobarometer. For all variables, the z-score is considered. Reported coefficients are shown with standard errors in parentheses. Regions considered at NUTS0 (country) level. * p $<$ 0.1, ** p $<$ 0.05, *** p $<$ 0.01.}
    \label{tab:regression_main_NUTS0}
\end{table}

\begin{table}[h]
    \begin{tabular}{lccccccc}
        \toprule
        \textbf{$ProEnv_{19}$} & (1) & (2) & (3) & (4) & (5) & (6) & (7)\\
        \midrule
        \textbf{$DenUsers$} & $0.499^{***}$ &  &  & $0.338^{***}$ & $0.283^{***}$ &  & $0.209^{**}$\\
         & (0.064) &  &  & (0.089) & (0.096) &  & (0.105)\\

        \textbf{$ES_{EDU}$} &  & $0.467^{***}$ &  & $0.228^{**}$ &  & $0.238^{***}$ & $0.164^{*}$\\
         &  & (0.065) &  & (0.089) &  & (0.085) & (0.092)\\

        \textbf{$ES_{PPS}$} &  &  & $0.499^{***}$ &  & $0.284^{***}$ & $0.338^{***}$ & $0.230^{**}$\\
         &  &  & (0.064) &  & (0.096) & (0.085) & (0.101)\\

        \midrule
        Number of regions & 185 & 185 & 185 & 185 & 185 & 185 & 185\\
        $R^2$ & 0.249 & 0.218 & 0.249 & 0.275 & 0.283 & 0.280 & 0.295\\
        \bottomrule
    \end{tabular}

    \caption{Regression results for the average number of pro-environmental actions ($ProEnv_{19}$). The main predictor is the number of users normalized by the regional population ($DenUsers$), while confounding factors include regional purchasing power per capita (Regional PPS) and the average regional education level in 2019, both computed from Eurostat. For all variables, we applied the logarithm and then computed the z-score. Reported coefficients are shown with standard errors in parentheses. Regions considered at NUTS2 level. * p $<$ 0.1, ** p $<$ 0.05, *** p $<$ 0.01. }
    \label{tab:regression_users_altES}
\end{table}

\begin{table}[h]
    \begin{tabular}{lccccccc}
        \toprule
        \textbf{$ProEnv_{19}$} & (1) & (2) & (3) & (4) & (5) & (6) & (7)\\
        \midrule
        \textbf{$DenUsers$} & $0.499^{***}$ &  &  & $0.400^{***}$ & $0.395^{***}$ &  & $0.354^{***}$\\
         & (0.064) &  &  & (0.063) & (0.060) &  & (0.060)\\

        \textbf{$EB_{EDU}$} &  & $0.445^{***}$ &  & $0.323^{***}$ &  & $0.284^{***}$ & $0.201^{***}$\\
         &  & (0.066) &  & (0.063) &  & (0.068) & (0.064)\\

        \textbf{$EB_{SOC}$} &  &  & $0.495^{***}$ &  & $0.390^{***}$ & $0.372^{***}$ & $0.314^{***}$\\
         &  &  & (0.064) &  & (0.060) & (0.068) & (0.063)\\

        \midrule
        Number of regions & 185 & 185 & 185 & 185 & 185 & 185 & 185\\
        $R^2$ & 0.249 & 0.198 & 0.245 & 0.343 & 0.390 & 0.311 & 0.422\\
        \bottomrule
    \end{tabular}

    \caption{Regression results for the average number of pro-environmental actions ($ProEnv_{19}$). The main predictor is the number of users normalized by the regional population ($DenUsers$), while confounding factors include the average number of years of full-time education ($EB_{EDU}$) and the average social class ($EB_{SOC}$), both computed from the Eurobarometer. For all variables, the z-score is considered. Reported coefficients are shown with standard errors in parentheses. Regions considered at NUTS2 level. * p $<$ 0.1, ** p $<$ 0.05, *** p $<$ 0.01.}
    \label{tab:regression_user_main}
\end{table}

\begin{table}[h]
    \begin{tabular}{lccccccc}
        \toprule
        \textbf{$ProEnv_{19}$} & (1) & (2) & (3) & (4) & (5) & (6) & (7)\\
        \midrule
        \textbf{$BelTweets$} & $0.501^{***}$ &  &  & $0.400^{***}$ & $0.396^{***}$ &  & $0.354^{***}$\\
         & (0.064) &  &  & (0.063) & (0.060) &  & (0.060)\\

        \textbf{$EB_{EDU}$} &  & $0.445^{***}$ &  & $0.318^{***}$ &  & $0.284^{***}$ & $0.197^{***}$\\
         &  & (0.066) &  & (0.063) &  & (0.068) & (0.065)\\

        \textbf{$EB_{SOC}$} &  &  & $0.495^{***}$ &  & $0.388^{***}$ & $0.372^{***}$ & $0.315^{***}$\\
         &  &  & (0.064) &  & (0.060) & (0.068) & (0.064)\\

        \midrule
        Number of regions & 185 & 185 & 185 & 185 & 185 & 185 & 185\\
        $R^2$ & 0.251 & 0.198 & 0.245 & 0.342 & 0.391 & 0.311 & 0.420\\
        \bottomrule
    \end{tabular}

    \caption{Regression results for the average number of pro-environmental actions ($ProEnv_{19}$). The main predictor is the number of believer tweets normalized by the regional population ($BelTweets$), while confounding factors include regional purchasing power per capita (Regional PPS) and the average regional education level in 2019, both computed from the Eurobarometer. For all variables, we applied the logarithm and then computed the z-score. Reported coefficients are shown with standard errors in parentheses. Regions considered at NUTS2 level. * p $<$ 0.1, ** p $<$ 0.05, *** p $<$ 0.01. }
    \label{tab:regression_main_bel}
\end{table}

\newpage
\newpage
\clearpage
\clearpage
\bibliography{bibliography_article}

@article{effrosynidis2022climate,
  title={The climate change Twitter dataset},
  author={Effrosynidis, Dimitrios and Karasakalidis, Alexandros I and Sylaios, Georgios and Arampatzis, Avi},
  journal={Expert Systems with Applications},
  volume={204},
  pages={117541},
  year={2022},
  publisher={Elsevier}
}

@article{effrosynidis2022exploring,
  title={Exploring climate change on Twitter using seven aspects: Stance, sentiment, aggressiveness, temperature, gender, topics, and disasters},
  author={Effrosynidis, Dimitrios and Sylaios, Georgios and Arampatzis, Avi},
  journal={Plos one},
  volume={17},
  number={9},
  pages={e0274213},
  year={2022},
  publisher={Public Library of Science San Francisco, CA USA}
}

@article{crispino2024people,
  title={Do people pay attention to climate change? Evidence from Italy},
  author={Crispino, Marta and Loberto, Michele},
  journal={Journal of Economic Behavior \& Organization},
  volume={219},
  pages={434--449},
  year={2024},
  publisher={Elsevier}
}

@article{nguyen2021social,
  title={Social incentive factors in interventions promoting sustainable behaviors: A meta-analysis},
  author={Nguyen-Van, Phu and Stenger, Anne and Tiet, Tuyen},
  journal={Plos one},
  volume={16},
  number={12},
  pages={e0260932},
  year={2021},
  publisher={Public Library of Science San Francisco, CA USA}
}

@article{aiello2022multidimensional,
  title={Multidimensional tie strength and economic development},
  author={Aiello, Luca Maria and Joglekar, Sagar and Quercia, Daniele},
  journal={Scientific Reports},
  volume={12},
  number={1},
  pages={22081},
  year={2022},
  publisher={Nature Publishing Group UK London}
}

@inproceedings{choi2020ten,
  title={Ten social dimensions of conversations and relationships},
  author={Choi, Minje and Aiello, Luca Maria and Varga, Kriszti{\'a}n Zsolt and Quercia, Daniele},
  booktitle={Proceedings of The Web Conference 2020},
  pages={1514--1525},
  year={2020}
}

@article{deri2018coloring,
  title={Coloring in the links: Capturing social ties as they are perceived},
  author={Deri, Sebastian and Rappaz, Jeremie and Aiello, Luca Maria and Quercia, Daniele},
  journal={Proceedings of the ACM on Human-Computer Interaction},
  volume={2},
  number={CSCW},
  pages={1--18},
  year={2018},
  publisher={ACM New York, NY, USA}
}

@misc{ZA7602,
author = "European Commission, Brussels",
title = "Eurobarometer 92.4 (2019)",
year = "2024",
howpublished = "GESIS, Cologne. ZA7602 Data file Version 2.0.0, https://doi.org/10.4232/1.14381",
doi = "10.4232/1.14381",
}

@inproceedings{pera2025extracting,
  title={Extracting Participation in Collective Action from Social Media},
  author={Pera, Arianna and Aiello, Luca Maria},
  booktitle={Proceedings of the International AAAI Conference on Web and Social Media},
  volume={19},
  pages={1530--1549},
  year={2025}
}

@article{sisco2017extreme,
  title={When do extreme weather events generate attention to climate change?},
  author={Sisco, Matthew R and Bosetti, Valentina and Weber, Elke U},
  journal={Climatic change},
  volume={143},
  number={1},
  pages={227--241},
  year={2017},
  publisher={Springer}
}

@article{kirilenko2015people,
  title={People as sensors: Mass media and local temperature influence climate change discussion on Twitter},
  author={Kirilenko, Andrei P and Molodtsova, Tatiana and Stepchenkova, Svetlana O},
  journal={Global Environmental Change},
  volume={30},
  pages={92--100},
  year={2015},
  publisher={Elsevier}
}

@article{kirilenko2014public,
  title={Public microblogging on climate change: One year of Twitter worldwide},
  author={Kirilenko, Andrei P and Stepchenkova, Svetlana O},
  journal={Global environmental change},
  volume={26},
  pages={171--182},
  year={2014},
  publisher={Elsevier}
}

@inproceedings{malik2015population,
  title={Population bias in geotagged tweets},
  author={Malik, Momin and Lamba, Hemank and Nakos, Constantine and Pfeffer, J{\"u}rgen},
  booktitle={proceedings of the international AAAI conference on web and social media},
  volume={9},
  number={4},
  pages={18--27},
  year={2015}
}

@article{longley2015geotemporal,
  title={The geotemporal demographics of Twitter usage},
  author={Longley, Paul A and Adnan, Muhammad and Lansley, Guy},
  journal={Environment and Planning A},
  volume={47},
  number={2},
  pages={465--484},
  year={2015},
  publisher={SAGE Publications Sage UK: London, England}
}

@article{pearce2019social,
  title={The social media life of climate change: Platforms, publics, and future imaginaries},
  author={Pearce, Warren and Niederer, Sabine and {\"O}zkula, Suay Melisa and S{\'a}nchez Querub{\'\i}n, Natalia},
  journal={Wiley interdisciplinary reviews: Climate change},
  volume={10},
  number={2},
  pages={e569},
  year={2019},
  publisher={Wiley Online Library}
}

@article{sultana2024systematic,
  title={A systematic review of the nexus between climate change and social media: present status, trends, and future challenges},
  author={Sultana, Bebe Chand and Prodhan, Md Tabiur Rahman and Alam, Edris and Sohel, Md Salman and Bari, ABM Mainul and Pal, Subodh Chandra and Islam, Md Kamrul and Islam, Abu Reza Md Towfiqul},
  journal={Frontiers in Communication},
  volume={9},
  pages={1301400},
  year={2024},
  publisher={Frontiers Media SA}
}

@article{williams2015network,
  title={Network analysis reveals open forums and echo chambers in social media discussions of climate change},
  author={Williams, Hywel TP and McMurray, James R and Kurz, Tim and Lambert, F Hugo},
  journal={Global environmental change},
  volume={32},
  pages={126--138},
  year={2015},
  publisher={Elsevier}
}

@article{gaupp2024climate,
  title={Climate Activism, Social Media and Behavioural Change: A Literature Review},
  author={Gaupp, Franziska and Eker, Sibel},
  year={2024},
  publisher={WP-24-007}
}

@article{pearson2016can,
  title={Can we tweet, post, and share our way to a more sustainable society? A review of the current contributions and future potential of\# socialmediaforsustainability},
  author={Pearson, Elissa and Tindle, Hayley and Ferguson, Monika and Ryan, Jillian and Litchfield, Carla},
  journal={Annual Review of Environment and Resources},
  volume={41},
  number={1},
  pages={363--397},
  year={2016},
  publisher={Annual Reviews}
}

@article{chatterjee2024communication,
  title={Communication, Climate Mitigation, and Behavior Change Interventions: Understanding Message Design and Digital Media Technologies},
  author={Chatterjee, Joyee Shairee and Thongprasert, Sirayuth and Some, Shreya},
  journal={Annual Review of Environment and Resources},
  volume={49},
  year={2024},
  publisher={Annual Reviews}
}

@article{fownes2018twitter,
  title={Twitter and climate change},
  author={Fownes, Jennifer R and Yu, Chao and Margolin, Drew B},
  journal={Sociology Compass},
  volume={12},
  number={6},
  pages={e12587},
  year={2018},
  publisher={Wiley Online Library}
}

@article{veltri2017climate,
  title={Climate change on Twitter: Content, media ecology and information sharing behaviour},
  author={Veltri, Giuseppe A and Atanasova, Dimitrinka},
  journal={Public understanding of science},
  volume={26},
  number={6},
  pages={721--737},
  year={2017},
  publisher={SAGE Publications Sage UK: London, England}
}

@article{loureiro2020sensing,
  title={Sensing climate change and energy issues: Sentiment and emotion analysis with social media in the UK and Spain},
  author={Loureiro, Maria L and All{\'o}, Maria},
  journal={Energy Policy},
  volume={143},
  pages={111490},
  year={2020},
  publisher={Elsevier}
}

@article{bennett2021mapping,
  title={Mapping climate discourse to climate opinion: An approach for augmenting surveys with social media to enhance understandings of climate opinion in the United States},
  author={Bennett, Jackson and Rachunok, Benjamin and Flage, Roger and Nateghi, Roshanak},
  journal={PloS one},
  volume={16},
  number={1},
  pages={e0245319},
  year={2021},
  publisher={Public Library of Science San Francisco, CA USA}
}

@article{kolic2025chambers,
  title={From chambers to echo chambers: quantifying polarization with a second-neighbor approach applied to Twitter’s climate discussion},
  author={Kolic, Blas and Aguirre-L{\'o}pez, Fabi{\'a}n and Hern{\'a}ndez-Williams, Sergio and Gardu{\~n}o-Hern{\'a}ndez, Guillermo},
  journal={Journal of Complex Networks},
  volume={13},
  number={4},
  pages={cnaf020},
  year={2025},
  publisher={Oxford University Press}
}

@article{sun2021estimating,
  title={Estimating local-scale domestic electricity energy consumption using demographic, nighttime light imagery and Twitter data},
  author={Sun, Yeran and Wang, Shaohua and Zhang, Xucai and Chan, Ting On and Wu, Wenjie},
  journal={Energy},
  volume={226},
  pages={120351},
  year={2021},
  publisher={Elsevier}
}

@article{crespo2023role,
  title={The role of social media activism in offline conservation attitudes and behaviors},
  author={Crespo, Yanitza Angely Cruz and Cruz, Shannon M},
  journal={Computers in Human Behavior},
  volume={147},
  pages={107858},
  year={2023},
  publisher={Elsevier}
}

@article{valori2012reconciling,
  title={Reconciling long-term cultural diversity and short-term collective social behavior},
  author={Valori, Luca and Picciolo, Francesco and Allansdottir, Agnes and Garlaschelli, Diego},
  journal={Proceedings of the National Academy of Sciences},
  volume={109},
  number={4},
  pages={1068--1073},
  year={2012},
  publisher={National Academy of Sciences}
}

@article{buabeanu2018ultrametricity,
  title={Ultrametricity increases the predictability of cultural dynamics},
  author={B{\u{a}}beanu, Alexandru-Ionu{\c{t}} and Vis, Jorinde van de and Garlaschelli, Diego},
  journal={New Journal of Physics},
  volume={20},
  number={10},
  pages={103026},
  year={2018},
  publisher={IOP Publishing}
}

@article{buabeanu2017signs,
  title={Signs of universality in the structure of culture},
  author={B{\u{a}}beanu, Alexandru-Ionu{\c{t}} and Talman, Leandros and Garlaschelli, Diego},
  journal={The European Physical Journal B},
  volume={90},
  number={12},
  pages={237},
  year={2017},
  publisher={Springer}
}

\end{document}